\pdfoutput=1

\documentclass[11pt]{article}

\usepackage[final]{coling}
\usepackage{amssymb}

\usepackage{times}
\usepackage{latexsym}
\usepackage{multirow}
\usepackage{colortbl}
\usepackage{diagbox}
\usepackage{tcolorbox}
\usepackage{amsmath}
\usepackage[T1]{fontenc}
\usepackage{algorithm}

\usepackage[utf8]{inputenc}

\usepackage{enumitem}
\usepackage{extarrows}
\usepackage{inconsolata}
\usepackage{booktabs}
\usepackage{graphicx}
\usepackage{subfigure}
\usepackage{algorithm}
\usepackage{algpseudocode}
\usepackage{makecell}
\newcommand{\smallsection}[1]{{\vspace{0.05in} \noindent \bf {#1.\hspace{5pt}}}}

\usepackage[nopatch=footnote]{microtype}
\usepackage{hyperref}
\usepackage{marvosym}

\title{\textsc{FunnelRAG}: A Coarse-to-Fine Progressive Retrieval Paradigm for RAG}

\author{Xinping Zhao$^{1}$,  Yan Zhong$^{2}$, Zetian Sun$^{1}$, Xinshuo Hu$^{1}$, \\  \textbf{Zhenyu Liu$^{1}$, Dongfang Li$^{1}$, Baotian Hu$^{1}$\textsuperscript{\Letter}\thanks{\textsuperscript{\Letter}Corresponding author.},  Min Zhang$^{1}$} \\ 
        $^{1}$Harbin Institute of Technology (Shenzhen), $^{2}$Peking University \\
        \texttt{\{zhaoxinping, 23S151141, 22S051034, 190110924\}@stu.hit.edu.cn}, \\ zhongyan@stu.pku.edu.cn, \texttt{\{lidongfang, hubaotian, zhangmin2021\}@hit.edu.cn}}
\makeatletter
\def\thanks#1{\protected@xdef\@thanks{\@thanks
        \protect\footnotetext{#1}}}
\makeatother

\begin{document}
\maketitle
\begin{abstract}
Retrieval-Augmented Generation (RAG) prevails in Large Language Models. It mainly consists of \textit{retrieval} and \textit{generation}.
The retrieval modules (\textit{a.k.a.} retrievers) aim to find useful information used to facilitate the generation modules (\textit{a.k.a.} generators).
As such, generators' performance largely depends on the effectiveness and efficiency of retrievers. 
However, the widely used retrieval paradigm remains flat. It treats retrieval procedures as a one-off deal with constant granularity.
Despite effectiveness, we argue that they suffer from two limitations: (1) \textbf{flat retrieval} exerts a significant burden on one retriever; (2) \textbf{constant granularity} limits the ceiling of retrieval performance.
In this work, we propose a progressive retrieval paradigm with coarse-to-fine granularity for RAG, termed \textsc{FunnelRAG}, so as to balance effectiveness and efficiency.
Specifically, \textsc{FunnelRAG} establishes a progressive retrieval pipeline by collaborating coarse-to-fine granularity, large-to-small quantity, and low-to-high capacity, which can relieve the burden on one retriever and also promote the ceiling of retrieval performance.
Extensive experiments manifest that \textsc{FunnelRAG} achieves comparable retrieval performance while the time overhead is reduced by nearly 40 percent.
\end{abstract}

\section{Introduction}
\label{sec:intro}
\begin{figure}[t]
    \centering
    \includegraphics[width=1.\linewidth]{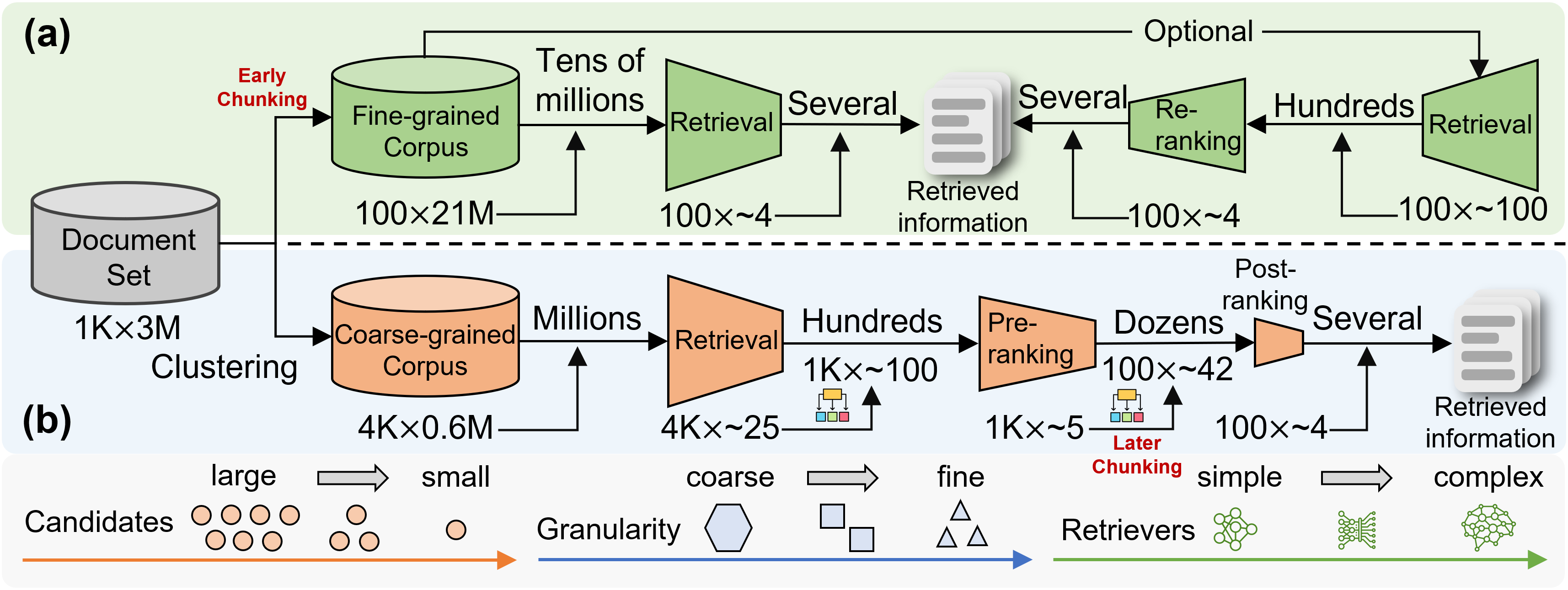}
    \caption{Comparison between \textbf{(a)} the flat retrieval and \textbf{(b)} the progressive retrieval paradigm, where \includegraphics[width=0.02\textwidth]{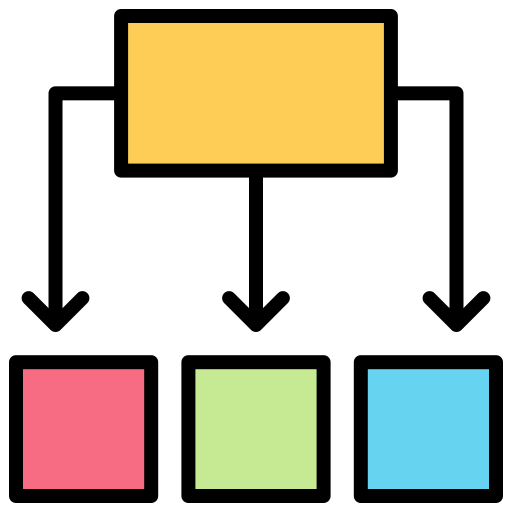} is the segmentation operation. \textsc{FunnelRAG} performs progressive retrieval from large to small quantity, from coarse to fine granularity, and with simple to complex retrievers, which balances effectiveness and efficiency.} 
    \label{fig:example}
\end{figure}

Retrieval-Augmented Generation (RAG) has been shown highly effective in enhancing Large Language Models (LLMs) \cite{DBLP:journals/corr/abs-2312-10997,DBLP:journals/corr/abs-2301-12652} and has been widely adopted in the industry, such as Microsoft's GraphRAG \cite{DBLP:journals/corr/abs-2404-16130}, Google's REALM \cite{DBLP:journals/corr/abs-2002-08909}, and Meta's RA-DIT \cite{DBLP:conf/iclr/Lin0CSL00KSLZY24}. Its effectiveness mainly comes from retrieving external non-parametric knowledge into LLMs to remedy their incomplete, incorrect, or outdated internal parametric knowledge \cite{DBLP:conf/emnlp/KarpukhinOMLWEC20,DBLP:journals/corr/abs-1911-03868}. 
The de facto RAG framework usually segments documents into short retrieval units, such as 100-word passages \cite{DBLP:journals/corr/abs-2406-15319}, resulting in a massive corpus with tens of millions of candidate units. Then, the retriever is tasked to find the ``needle'' (\textit{i.e.,} the golden retrieval units) from the ``haystack'' (\textit{i.e.,} the enormous candidate corpus) \cite{DBLP:journals/corr/abs-2406-13121,GregNeedle}. Finally, the retrieved units serve as the input context to the generator to facilitate generation. 
Its working flow is shown in Figure \ref{fig:example}(a). Wikipedia dump is used as the non-parametric knowledge source \cite{DBLP:conf/nips/LewisPPPKGKLYR020}, where each document is segmented into 100-word chunks, resulting in a total of 21M short passages. Then, the retriever needs to seek through a vast number of 21M candidates to get several potentially valuable passages, such as four. 
Despite effectiveness, the existing retrieval paradigms still suffer from two major\, limitations:
\begin{table*}[h]
  \centering
  \footnotesize
  \tabcolsep=0.185cm
  \renewcommand\arraystretch{1.2}
  \begin{tabular}{ccccc}
    \toprule
    \multicolumn{1}{l}{\textbf{Retrieval Paradigm}} & 
    \multicolumn{1}{c}{\textbf{Retrieval Unit}} & 
    \multicolumn{1}{c}{\textbf{Time to Chunk}} & 
    \multicolumn{1}{c}{\textbf{Corpus Size}} &
     \multicolumn{1}{c}{\textbf{Unit Size}}\\
    \midrule
    \multirow{1}{*}{{Flat Retrieval}} & \multirow{1}{*}{\makecell{Passage $\rightarrow$ Passage (Optional)}} & Early Chunking & 21M & 100 \\
    \multirow{1}{*}{{Progressive Retrieval}} & \multirow{1}{*}{\makecell{Clustered Document$\rightarrow$ Document $\rightarrow$ Passage }} & Later Chunking & 600K & 4K \\
    \bottomrule
  \end{tabular}
  \caption{\centering{Key feature comparison between the prevailing flat retrieval and the proposed progressive retrieval.}}
  \label{tab:retrieval_statistics}
\end{table*}
\begin{itemize}[leftmargin=*]
    \item \textbf{Flat Retrieval.} Most RAG frameworks approach the retrieval stage as a one-off deal, where the retriever is requested to take tens of millions of candidates as input and to find the golden retrieval units at a sitting. However, these practices inflict a heavy burden on one retriever, which makes the retrieval stage less effective and efficient, since it is very tough and computationally complex to find golden units immediately from such an\, enormous\, candidate\, pool\, (\textit{e.g.,} 21M). 
    
    \item \textbf{Constant Granularity.} Some RAG frameworks draw inspiration from recommender systems \cite{DBLP:journals/corr/abs-2007-16122,DBLP:conf/recsys/CovingtonAS16,DBLP:conf/kdd/GrbovicC18}, following a multi-stage cascade architecture, where candidates are extracted through \textit{retrieval} and \textit{reranking}, as shown in Figure \ref{fig:example}(a). However, these practices directly introduce recommender systems' experience into RAG and neglect the characteristics of RAG. Each entry in recommender systems is usually inseparable, while entries in RAG are usually separable. Unfortunately, existing RAG methods commonly treat entries as constant-grained retrieval units, ultimately restricting\, performance.
\end{itemize}
The above issues inflict a heavy burden on one retriever and neglect the subdivisible characteristic of entries in RAG. In this work, we propose a progressive retrieval paradigm with coarse-to-fine granularity for RAG, termed \textsc{FunnelRAG}. 
Specifically, it progressively reduces the scale of candidate entries, refines the granularity of retrieval units, and increases the level of retrievers' capacity, whose working flow is shown in Figure \ref{fig:example}(b). Particularly, there are two important designs in \textsc{FunnelRAG}:
\begin{itemize}[leftmargin=*]
    \item \textbf{Progressive Retrieval.} Different from flat retrieval, we progressively reduce the scale of candidates (\textit{e.g.,} millions (0.6M) $\rightarrow$ hundreds (100) $\rightarrow$ dozens (42)) and increase the level of retrievers' capacity (\textit{e.g.,} sparse retrievers (SR)$\rightarrow$ dense retrievers (DR)$\rightarrow$small language models (SLM)), to make a better balance of effectiveness and efficiency. 
    This design enables load balancing and improves retrieval accuracy by using mixed-capacity retrievers. To improve the retrieval accuracy of the entire pipeline, we train SLM with retrieval-augmented fine-tuning to gain retrieval capacity. Then, we further distill the aggregated retrieval signals from SLM to DR so that DR could align\, with\, language models' preferences. 

    \item \textbf{Coarse-to-Fine Granularity.} To complement mixed-capacity retrievers with each other, it is necessary to segment coarse-grained units into fine-grained ones, as the high-capacity retrievers (\textit{e.g.,} DR) perform relatively poor than low-capacity ones (\textit{e.g.,} SR) for coarse-grained units \cite{DBLP:journals/corr/abs-2402-03216}. As such, we first construct coarse-grained units with approximate 4K tokens by clustering multiple related documents before retrieval, which can considerably reduce the corpus size (\textit{e.g.,} 21M$\rightarrow$0.6M). 
    Then, coarse-grained units are segmented into document-level units (\textit{e.g.,} 4K$\rightarrow$1K) before pre-ranking. Finally, we segment document-level units into passage-level units (\textit{e.g.,} 1K$\rightarrow$100) before post-ranking. These three groups of varied granularity units are sequentially fed into SR, DR, and SLM to locate the golden units with high accuracy and low cost.
\end{itemize}
These two novel designs jointly contribute to the considerable improvement of retrieval accuracy and efficiency in open-domain question answering (QA), such as Natural Question (NQ) \cite{DBLP:journals/tacl/KwiatkowskiPRCP19} and Trivia QA (TQA) \cite{DBLP:conf/acl/JoshiCWZ17}.
Table \ref{tab:retrieval_statistics} shows the key differences between the progressive retrieval and the flat one.
\textsc{FunnelRAG} features \textbf{(1)} Coarse-to-fine granularity which balances load and accuracy; \textbf{(2)} Later chunking which perceives the contextual information of retrieval units well\footnote{Flat retrieval reranks passages that are not contextual integrity. However, our progressive retrieval perceives contextual information well in the post-ranking stage, since these passages are derived from documents with contextual integrity.}; 
\textbf{(3)} Compressed corpus (30x smaller from 21M to 600K) which reduces the burden on the retrieval stage; and \textbf{(4)} Long retrieval unit which improves answer recall to the full extent. More details can be found in \S \ref{sec:method}.
The main contributions of this work are summarized as three-folds: \textbf{(1)} This work highlights the issues commonly encountered in real-world RAG systems while overlooked in the existing studies, \textit{i.e.,} the flat retrieval and constant granularity issues. 
\textbf{(2)} This work proposes \textsc{FunnelRAG}, a coarse-to-fine progressive retrieval paradigm for RAG, satisfying the three properties of being time-saving, fine-grained, and contextual-integrity. 
\textbf{(3)} Extensive experiments demonstrate that \textsc{FunnelRAG} can considerably reduce time overhead while the retrieval performance is comparable or even better in comparison\, with\, existing\, retrieval\, paradigms.

\section{Preliminaries}
\subsection{General Retrieval Paradigm for RAG}
In retrieval-augmented generation, we are given a query $q$ and a document set $\mathcal{D}=\{d_1, d_2, ..., d_{|\mathcal{D}|}\}$, where each document will be usually segmented into passage-level units, which results in a fine-grained corpus $\mathcal{P}=\{p_1, p_2, ..., p_{|\mathcal{P}|}\}$ consisting of millions of short retrieval units. The de facto retrieval paradigm comprises two stages: \textit{retrieval} and \textit{reranking}. 
The retrieval stage mostly employs a dense retriever, such as the Dense Passage Retriever \cite{DBLP:conf/emnlp/KarpukhinOMLWEC20}, which projects each passage $p \in \mathcal{P}$ to an embedding $\mathbf{E}(p)$ and projects the query $q$ to an embedding $\mathbf{E}(q)$\footnote{Note that the embedding encoder $\mathbf{E}(\cdot)$ used for encoding queries and passages may not be the same, refer to \cite{DBLP:conf/emnlp/KarpukhinOMLWEC20}. For\, simplicity, we\, do make no\, distinction\, here.}. Then, the top-$k$ relevant passages for query $q$ are retrieved based on the query-passage embedding similarity, which is often computed by the dot product $\mathbf{E}(q)^\mathrm{T}\mathbf{E}(p)$.
After that, the reranking stage employs a higher capacity reranker (such as RankGPT \cite{DBLP:conf/emnlp/0001YMWRCYR23}) to filter out irrelevant passages. In the last, the top-$n$ passages are fed into the context\, of\, the\, generator to\, facilitate\, its\, generation.

\begin{figure}[t]
    \centering  
    \subfigure[NQ dataset.]{
        \includegraphics[width=0.493\linewidth]{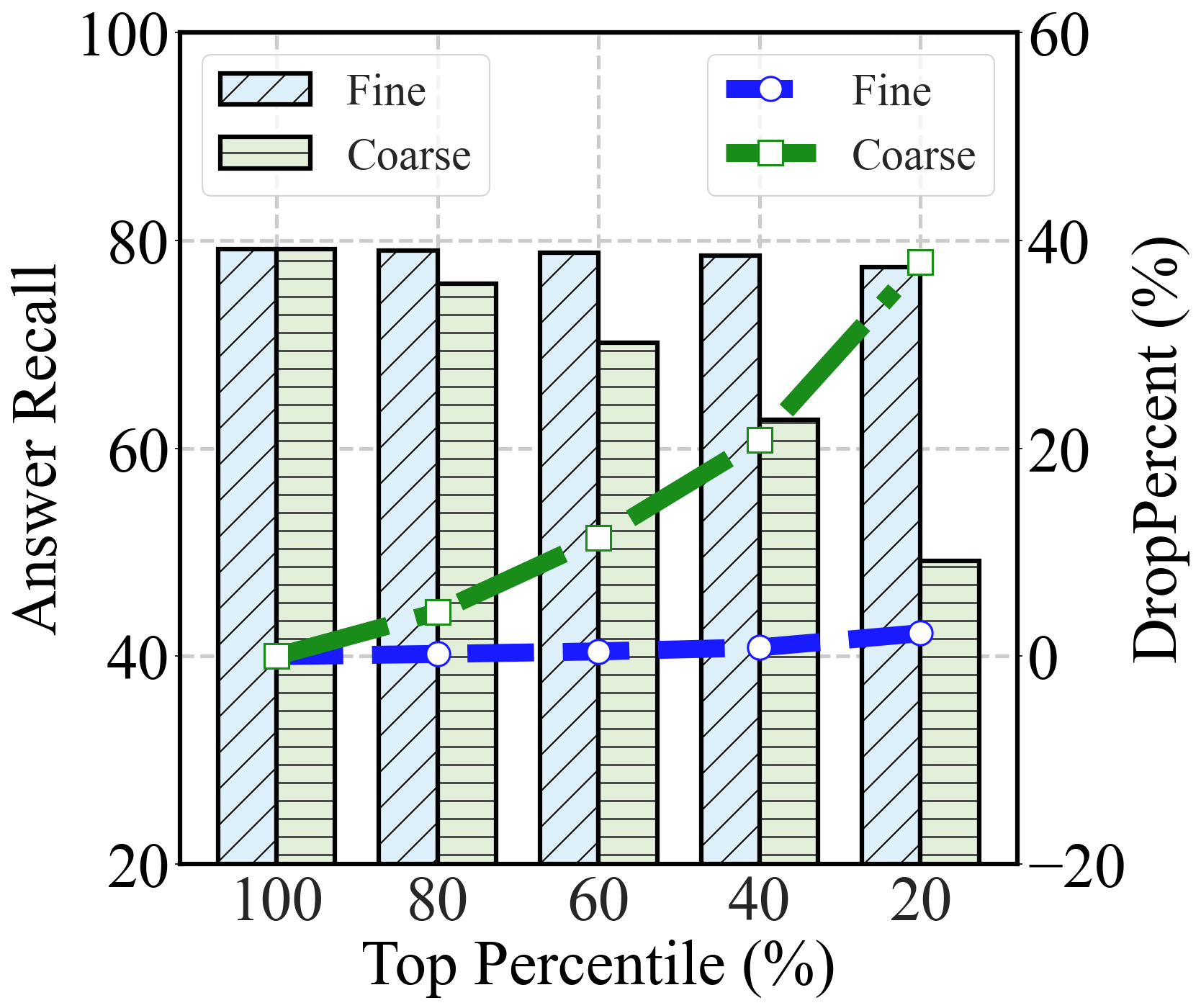}\label{fig:nq_coarse_to_fine}}
    \subfigure[TQA dataset.]{
        \includegraphics[width=0.469\linewidth]{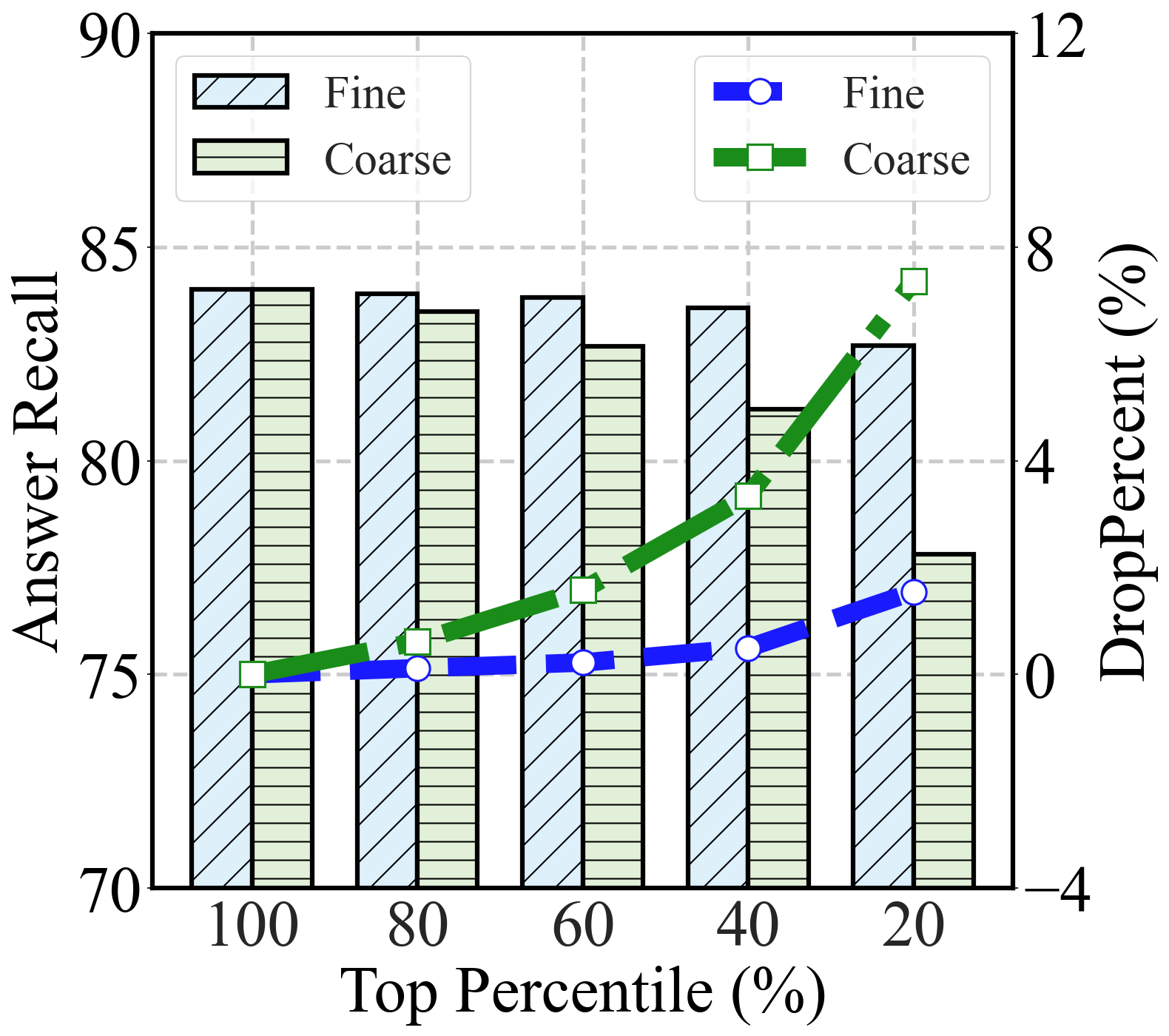}\label{fig:tqa_coarse_to_fine}}
    \caption{AR \textit{w.r.t.} coarse- and fine-grained retrieval. The bar denotes AR, while the line denotes the percentage of performance degradation compared to the cutoff position of 100\%. The X-axis represents the percentage of units retrieved. Under the same percentile, the number of tokens retrieved by `Fine' and `Coarse' is equal.}
    \label{fig:coarse_to_fine}
\end{figure}
\begin{figure}[t]
    \centering  
    \subfigure[NQ dataset.]{
        \includegraphics[width=1.0\linewidth]{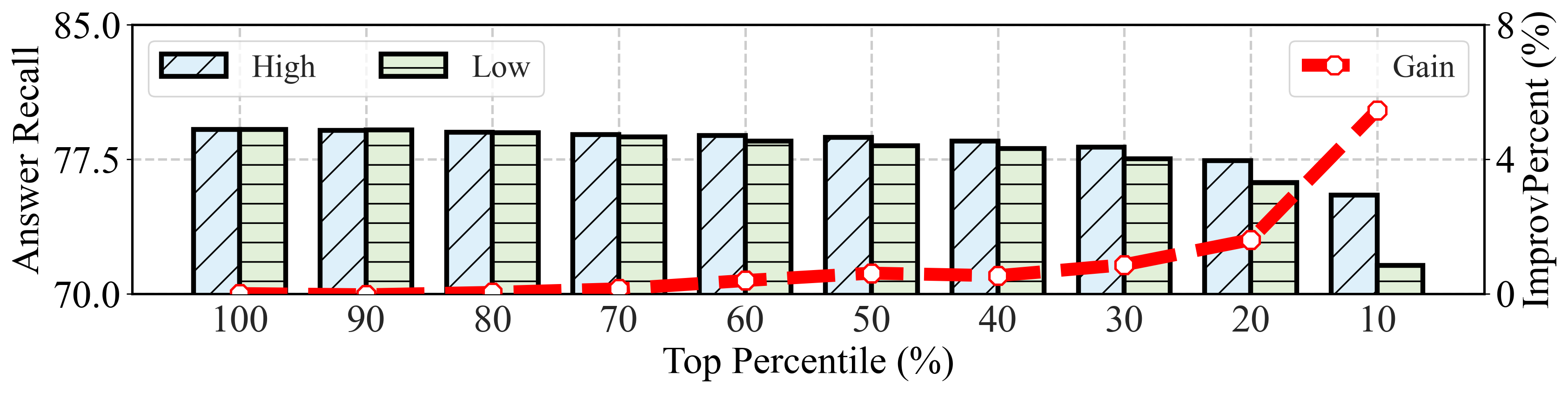}\label{fig:nq_high_to_low}}
    \subfigure[TQA dataset.]{
        \includegraphics[width=1.0\linewidth]{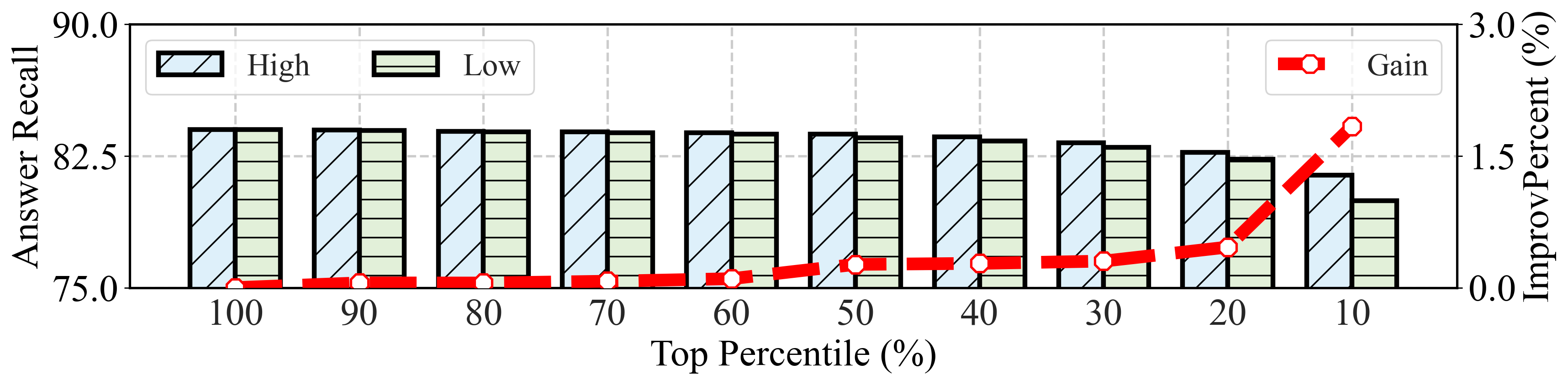}\label{fig:tqa_high_to_low}}
    \caption{Answer Recall \textit{w.r.t.} high- and low-capacity retrievers. The line denotes the percentage of performance improvement compared to\, the\, low-capacity\, retriever.}
    \label{fig:high_to_low}
\end{figure}
\subsection{Experiment for Assumption Validation}
As stated in Section \ref{sec:intro}, we emphasize the significance of (i) reducing the scale of candidates, (ii) refining the granularity of units, and (iii) increasing the level of retrievers' capacity, throughout the retrieval paradigm. Here, we conduct empirical studies to verify {claims} (ii) and (iii), as it is obvious that claim (i) will lead to better retrieval accuracy\footnote{Assuming a fixed number of golden units that exists in candidates, increasing the scale of candidates reduces the signal-to-noise ratio, inevitable degrading retrieval accuracy.} \cite{DBLP:journals/corr/abs-2404-07220}. We aim to evaluate whether these two claims can contribute to better retrieval performance, so as to validate that the progressive retrieval paradigm can balance effectiveness and efficiency well.
To empirically verify these claims, we construct two synthetic datasets for NQ and TQA. Specifically, for each query in NQ and TQA, we retrieve the top-10 relevant clustered documents from the coarse-grained corpus\footnote{Referring to Section \ref{sec:retrieval_stage} and Algorithm \ref{alg:cluster_documents} for more technical details about how the clustered\, documents\, formulated.} using BM25 \cite{DBLP:journals/ftir/RobertsonZ09}, where each one has approximate 4K tokens. Then, we segment each clustered document into document-level units with about 1K tokens, striking the coarse-fine contrast. On the other hand, we adopt bge-reranker-v2-m3 \cite{DBLP:journals/corr/abs-2402-03216} and BM25 to make a contrast between high- and low-capacity retrievers:
\begin{itemize}[leftmargin=*]
    \item \textbf{Coarse-Fine Contrast.} To simulate this scenario, we fix the retriever as bge-reranker-v2-m3 and use it to rerank coarse-grained (\textit{i.e.,} clustered documents) and fine-grained units (\textit{i.e.,} documents), respectively. We report the answer recall (AR) performance on the NQ and TQA\footnote{Answer Recall (AR) measures the recall of the answer string in all the retrieved units. We employ it as the retrieval metric, referring to Appendix 
\ref{app:ar} for more technical details.}, which are presented in Figures \ref{fig:nq_coarse_to_fine} and \ref{fig:tqa_coarse_to_fine}, respectively. 
    From the results, the answer recall with fine-grained retrieval substantially outperforms that with coarse-grained one. Additionally, the performance degradation with fine-grained retrieval is significantly slower than that with coarse-grained one. For example, on the NQ dataset, fine-grained retrieval only drops 2.25\% of its original performance, while coarse-grained retrieval drops 37.93\%, when the cutoff position is top 20\%. Given the above, it is necessary and valuable to segment coarse-grained units into fine-grained units along progressive retrieval stages in order for\, better\, retrieval performance.

    \item \textbf{High-Low Contrast.} To simulate this scenario, we fix the granularity of retrieval units as fine-grained ones and employ bge-reranker-v2-m3 and BM25 to rerank them, respectively. The experimental results are presented in Figure \ref{fig:nq_high_to_low} and \ref{fig:tqa_high_to_low}, respectively. From the results, we observe that the answer recall with the high-capacity retriever is consistently higher than that with the low-capacity one. In particular, the gain brought by the high-capacity retriever generally increases as the cutoff position decreases. 
    For example, on the NQ dataset, the retrieval performance improvement of `High' over `Low' is 5.45\% and 0.60\% in terms of cutoff@10\% and cutoff@20\%. These observations indicate that using high-capacity retrievers in later stages can retain the golden retrieval units to a large extent.
\end{itemize}

\section{Methodology}
\label{sec:method}
\begin{figure*}[t]
    \centering
    \includegraphics[width=1.0 \linewidth]{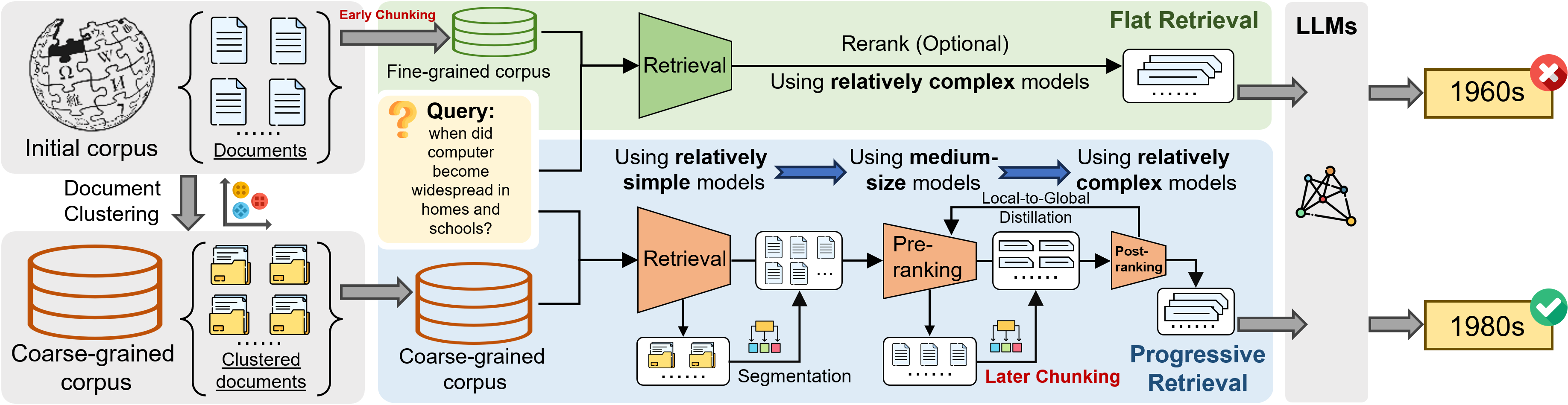}
    \caption{The overall system framework of \textsc{FunnelRAG}. The upper layer illustrates the working flow of the \textbf{\textcolor[RGB]{56,87,35}{flat retrieval paradigm}}, while\, the bottom\, layer\, illustrates the\, working flow\, of our \textbf{\textcolor[RGB]{31,78,121}{progressive retrieval paradigm}}.} 
    \label{fig:framework}
\end{figure*}

The overall framework of \textsc{FunnelRAG} is shown in Figure \ref{fig:framework}. We first introduce three progressive retrieval stages from coarse to fine (\S \ref{sec:progressive_retrieval}) and then describe how to distill aggregated signals from the succeeding retriever\, to the\, preceding\, one\, (\S \ref{sec:aggregated_signal}).

\subsection{Coarse-to-Fine Progressive Retrieval}
\label{sec:progressive_retrieval}
To better balance effectiveness and efficiency, we propose a progressive retrieval paradigm, which enables load balancing and improves retrieval accuracy by collaborating mixed-capacity retrievers, refining retrieval granularity, and reducing candidate scale.
It can be formulated into three stages: (1) Retrieval, which usually uses relatively simple bi-encoder models to retrieve coarse-grained units; (2) Pre-ranking, which adopts cross-encoder models to rank the previously retrieved units; and (3) Post-ranking, which employs relatively complex list-wise models to rank these fine-grained units. We provide an intuitive example in Appendix \ref{appen:case_study}.
\subsubsection{Retrieval Stage}
\label{sec:retrieval_stage}
This stage plays a role in improving answer recall (AR) to the full extent, without considering the granularity of the retrieval unit. Toward this end, we propose to convert the document set $\mathcal{D}$ into long retrieval (coarse-grained) units, which is proven to be effective in improving AR \cite{DBLP:journals/corr/abs-2406-15319}. 
Specifically, we cluster documents in $\mathcal{D}$ based on their relationships, \textit{i.e.,} hyperlinks embedded within each document, and finally results in\, a\, coarse-grained\, corpus\, $\mathcal{C}=\{c_1, c_2, ..., c_{|\mathcal{C}|}\}$. 
The clustering algorithm is presented in Appendix \ref{appen:cluster_document_algorithm} in Algorithm \ref{alg:cluster_documents}, where we draw inspiration from \cite{DBLP:journals/corr/abs-2406-15319} but make several major modifications. 
Specially, we sort documents by their local cluster coefficient from high to low in Line 1, so that documents with highly relevant will be clustered first. 
Furthermore, we measure the closeness centrality between document $d$ and its related clusters $\mathcal{R}$ and sort them from high to low in Line 10, so that the most related cluster will be merged first.  Each cluster $c$ in $\mathcal{C}$ is a list of documents related to each other. 
We set the max cluster size $S$ as 4K tokens because the experimental results (\S\ref{sec:study_of_funnelrag}) demonstrate that overlength retrieval units do not bring much benefits to answer\, recall.
Currently, it is challenging and underperforming for dense retrievers to handle long context, even with the SOTA dense retrievers, \textit{e.g,} BGE-M3 \cite{DBLP:journals/corr/abs-2402-03216}. 
In LongRAG \cite{DBLP:journals/corr/abs-2406-15319}, they handle long context by maximizing the scores of all chunks within each long retrieval unit, where they segment each long unit into 512-token chunks.
However, this practice has no essential difference with direct retrieval of short units and is computation-consuming. As the goal of this stage is to improve AR regardless of granularity, we use sparse retrievers to handle long context, which shows more effective than dense retrievers in long-doc retrieval \cite{DBLP:journals/corr/abs-2402-03216}. 
Without loss of generality, we adopt the representative sparse retriever in this work, \textit{i.e.,} BM25 \cite{DBLP:journals/ftir/RobertsonZ09}, whose computational cost is far below dense retrievers. 
Lastly, we use BM25 to retrieve the top-$K$ relevant long units $\mathcal{C}_{{l}} = \{c_1^l, c_2^l, ..., c_K^l\}$. We set $K$ as a proper value (\textit{e.g.,} 80) to achieve relatively high AR and reduce the computational burden\, of the succeeding stages.

\subsubsection{Pre-ranking Stage}
This stage plays a role in finding the ``needle'' (\textit{i.e.,} the document containing the answer string) from the ``needle case'' (\textit{i.e.,} the retrieved clusters) rather than the ``haystack'' (millions in most cases). 
The corpus size is the key difference between the existing retrieval paradigm and ours, where the corpus size of ours is hundreds, making it easy to find the ``needle''. Formally, we first segment each cluster $c^{l}$ in $\mathcal{C}_{{l}}$ into document-level units $d$. And then, we utilize a cross-encoder model, denoted as $f(\cdot)$, to measure their\, pre-ranking\, scores\, to\, the query\, $q$: 
\begin{equation}
    S^{{pre}} =f(q, d).
\end{equation}
Given pre-ranking scores, we retrieve top-$N$ documents closest to the query $\mathcal{D}_{m}=\{d_1^{m}, d_2^{m}, ..., d_{N}^{m}\}$. 
Figure \ref{fig:coarse_to_fine} shows the performance comparison between fine-grained and coarse-grained units. The results indicate that the finer retrieval units can significantly reduce the loss of retrieval performance. 
\subsubsection{Post-ranking Stage}
\label{sec:post_ranking_stage}
This stage plays a role in identifying fine-grained units that align with language models' preferences. We segment each document into passage-level units $p$ in this relatively late stage. 
Referring to \cite{JinaLateChunking}, we term this technique as ``\textbf{Later Chunking}'', which retrieves long units in the previous stages to preserve contextual semantics well and then segments long units into fine-grained units in the later stages for better retrieval applications. 
After chunking, we obtain $M$ passages $\{p_1^f, p_2^f, ..., p_M^f\}$. Then, a practical question naturally arises: \textit{How do efficiently post-rank these fine-grained units to align with language models' preferences?} 
A straightforward way is to serve LLMs as agents to post-rank passages with their preferences, \textit{e.g.,} RankGPT \cite{DBLP:conf/emnlp/0001YMWRCYR23} and RankVicuna \cite{DBLP:journals/corr/abs-2309-15088}. 
Nevertheless, directly using LLMs as rankers inevitably inflicts a significant computational burden on the post-ranking stage, making it\, very\, difficult\, to\, deploy\, in\, production. 
Instead of heuristically instructing LLMs to post-rank, our post-ranker builds upon FiD (Fusion-in-Decoder) \cite{DBLP:conf/eacl/IzacardG21}, a retrieval-augmented encoder-decoder language model. Specifically, each passage $p$ is paired with the query $q$ to be processed separately by the T5 \cite{DBLP:journals/jmlr/RaffelSRLNMZLL20} encoder. 
Subsequently, the encoded representations are concatenated along the sequence dimension. Lastly, the T5 \cite{DBLP:journals/jmlr/RaffelSRLNMZLL20} decoder attends to all of the parts simultaneously:
\begin{equation}
\label{eq:fid}
    \text{FiD}(q, \mathcal{P}_f) = \text{Dec}\big(\mathbf{H}_{q, p_1^f}\|, ...,\| \mathbf{H}_{q, p_M^f}\big),
\end{equation}
where $\mathbf{H}_{q, p^f}=\text{Enc}(q \oplus p^f)$; $\oplus$ and $\|$ are the concatenation operator. By independently processing each ($q$, $p^f$) pair, the encoder computes self-attention within one passage at a time. 
As such, the computational cost scales linearly with the number of passages, making the post-ranking stage highly efficient. 
Furthermore, the decoder's cross-attention scores have shown to be remarkably effective in estimating the relative importance of the retrieval-augmented passage from the language models' preferences
\cite{DBLP:conf/iclr/IzacardG21,DBLP:conf/acl/YuXY023,DBLP:journals/jmlr/IzacardLLHPSDJRG23,DBLP:conf/naacl/ShiMYS0LZY24}.
Given that, we average FiD cross-attention scores corresponding to each ($q$, $p^f$) pair from the first decoder token\footnote{Here, we take the score of the first token, as it leads to satisfactory performance in general. 
More analysis about attention aggregation schema can be found in Appendix \ref{appen:study_of_attention_aggregation_schema}.} over all the layers, all the heads, as well as all the representative tokens within $q\oplus p^f$: 
\begin{equation}
\label{equ:average_fid_score}
    S^{{post}} = \frac{1}{{ln} \times {lh} \times {lr}} \sum_{i=0}^{{ln}}\sum_{j=0}^{{lh}}\sum_{k=0}^{{lr}} \alpha_{0,i,j,\mathbf{r}[k]},
\end{equation}
where ${ln}$, ${lh}$, and ${lr}$ denote the number of the layers, heads, and representative tokens, respectively; $\alpha_{0,i,j,\mathbf{r}[k]}$ represents the score of the first decoder token in terms of the $i$-th layer, $j$-th head, and $k$-th representative token;
$\mathbf{r}[k]$ lookups the index of $k$-th representative token. We select ${lr}$ tokens with the highest scores to serve as representative ones\footnote{Note that we do not consider query tokens in $q$ when selecting representative tokens, refer to the ablation study 
in Appendix \ref{appen:ablation_of_query_tokens} for more details. More technical details about representative token selection can be found in Appendix \ref{appen:representative_token_selection}.}. 
The cross-attention mechanism is provided in Appendix \ref{appen:cross_attention}. Before estimating the relative importance score, we train FiD to predict the answer with retrieval-augmented passages, so that FiD can learn to look for cues, referring to Appendix \ref{app:fid} for more details. After that, we obtain the post-ranking score and treat these passages with scores ranked in the top-$H$ as oracle passages $\mathcal{P}_f=\{p_1^f, p_2^f, ..., p_H^f\}$. The oracle passages $\mathcal{P}_f$ will be fed into generators to\, facilitate\, generation.
\subsection{Distillation with Aggregated Signals}
\label{sec:aggregated_signal}
Ideally, we would like the retrieval units containing oracle passages to be ranked at the forefront by the preceding retriever, so that the succeeding one can reach out to oracle passages. 
To this end, we propose to distill retrieval knowledge from the succeeding to the preceding, termed local-to-global (L2G) distillation\footnote{In this work, we only distill retrieval knowledge from the post-ranking to pre-ranking stage, as we use sparse retrieval (BM25 \cite{DBLP:journals/ftir/RobertsonZ09}) in the retrieval stage.}. While distilling, an issue emerges: {the granularity of retrieval units between sequentially neighboring stages is different, hindering distillation.} 
Given that, we propose aggregating retrieval scores from local to global. By doing so, we can fill the granularity gap between two neighboring stages. Formally, we aggregate retrieval scores as:
\begin{equation}
\begin{split}
    S^{post  \rightarrow pre}_{d}  =  \alpha \times \max_{p \in d} \left\{S^{{post}} _{p}\right\} +& \\
      (1-\alpha) \times \mathop{\text{mean}}\limits_{p \in d} \left\{S^{{post}}_{p}\right\},& 
\end{split}
\end{equation}
where $\alpha \in [0, 1]$ controls the strength of `\textbf{max}' and `\textbf{mean}' parts. When $\alpha=1$, the score focuses on the most important passage in the document; when $\alpha=0$, the score focuses on the average importance level of all passages in the document. 
We annotate documents with $\text{Top-}\mathcal{K}$ aggregated scores as positive. Inspired by \cite{DBLP:conf/acl/YuXY023}, we annotate the ones $\mathcal{D}^{h+}$ that \textbf{h}it the ground truth as positive to make the training process more robust:
\begin{equation}
\begin{split}
    \mathcal{D}^+ &= \mathcal{D}^{h+} \cup 
    \mathop{\text{Top-}\mathcal{K}}\limits_{S^{post  \rightarrow pre}_{d}} \mathcal{D}_m, \\
    \mathcal{D}^- &=  \mathcal{D}_m \backslash \mathcal{D}^+,
\end{split}
\end{equation}
where the remaining documents (\textit{i.e.,} $\mathcal{D}_m \backslash \mathcal{D}^+$) serves as negative ones. 
Following ANCE \cite{DBLP:conf/iclr/XiongXLTLBAO21}, we adopt the pairwise Bayesian Personalized Ranking (BPR) loss \cite{DBLP:conf/uai/RendleFGS09}, enforcing the match score between the query and positive documents to be higher than negative ones:
\begin{equation}
    \mathcal{L} = \sum_{q}\sum_{d^+\in\mathcal{D}^+}\sum_{d^-\in\mathcal{D}^-} -\log\sigma(S^{{pre}}_{d^+}-S^{{pre}}_{d^-}),
\end{equation}
where $\sigma(\cdot)$ denotes the sigmoid function; $S^{{pre}}_{d^+}$ and $S^{{pre}}_{d^-}$ denote\, ${f}(q, d^+)$\, and\, ${f}(q, d^-)$,\, respectively. 
\begin{table*}[!ht]
\centering
\footnotesize
\tabcolsep=0.125cm
\renewcommand\arraystretch{1.2}
\begin{tabular}{ccllllc}
\toprule
    \multirow{2}{*}[-0.5ex]{\textbf{Datasets}} &
    \multirow{2}{*}[-0.5ex]{\textbf{Retrieval Paradigm}} &
    \multicolumn{3}{c}{\textbf{Retrieval Stages}} &
    \multicolumn{1}{c}{\multirow{2}{*}[-0.4ex]{\textbf{\begin{tabular}[c]{@{}c@{}}Time Cost \\ (T)\end{tabular}}}} &
    \multirow{2}{*}[-0.4ex]{\textbf{\begin{tabular}[c]{@{}c@{}}Answer Recall \\ (AR)\end{tabular}}} \\ 
    \cmidrule(lr){3-5}
    &  &  \multicolumn{1}{c}{\centering Retrieval} & \multicolumn{1}{c}{Pre/Re-ranking} & \multicolumn{1}{c}{Post-ranking} &  &  \\ 
    \midrule

    & \multirow{2}{*}{Flat Retrieval} &   21M $\rightarrow$ 4p  &  N/A  &  N/A   &  4.90 (4.90+N/A+N/A) & 72.91 \\
    & & 21M $\rightarrow$ 400p & 400p $\rightarrow$ 4p &  N/A &  5.25 (4.90+0.35+N/A) & 75.90 \\
    \rowcolor{gray!12}
    \cellcolor{white!20} &  & 600K $\rightarrow$ 4c & N/A  &  N/A  &  0.00 (0.00+N/A+N/A) &  71.69 \\
    
    \rowcolor{gray!12}
   \cellcolor{white!20} &  & 600K $\rightarrow$ 20c & 20c $\xrightarrow[]{\sim \text{180d}}$ 4d  &  N/A  &  0.49 (0.00+0.49+N/A) & 74.57  \\
    \rowcolor{gray!12}
    \cellcolor{white!20} \multirow{-5}{*}{\centering {NQ}} & \multirow{-3}{*}[0.5ex]{Progressive Retrieval} & 600K $\rightarrow$ 80c & 80c $\xrightarrow[]{\sim \text{740d}}$ 8d & 8d $\xrightarrow[]{\sim \text{50p}}$ 4p & 2.97 (0.00+2.20+0.77) & 75.43\\
    
    \midrule

     & \multirow{2}{*}{Flat Retrieval}  &   21M $\rightarrow$ 4p  &  N/A  &  N/A   &  5.02 (5.02+N/A+N/A) & 75.22  \\
    &  & 21M $\rightarrow$ 400p  &  400p $\rightarrow$ 4p  &  N/A   &  5.41 (5.02+0.39+N/A) & 81.29 \\
    \rowcolor{gray!12}
    \cellcolor{white!20} &  & 600K $\rightarrow$ 4c & N/A  &  N/A  &  0.00 (0.00+N/A+N/A) &  78.94 \\
    \rowcolor{gray!12}
    \cellcolor{white!20} &  & 600K $\rightarrow$ 20c & 20c $\xrightarrow[]{\sim \text{200d}}$ 4d  &  N/A  &  0.60 (0.00+0.60+N/A) &  80.00 \\
    \rowcolor{gray!12}
    \cellcolor{white!20} \multirow{-5}{*} {\centering {TQA}} & \multirow{-3}{*}[0.5ex]{\centering  Progressive Retrieval} & 600K $\rightarrow$ 80c & 80c $\xrightarrow[]{\sim \text{800d}}$ 12d & 12d $\xrightarrow[]{\sim \text{65p}}$ 4p & 3.47 (0.00+2.52+0.95) & 80.69 \\

\bottomrule
\end{tabular}
\caption{Retrieval performance comparison \textit{w.r.t.} time cost and answer recall on NQ and TQA datasets, where `c', `d', and `p' denote clustered documents, documents, and passages, respectively. The value on the arrow ($\rightarrow$) indicates the number of fine-grained candidates after segmenting coarse-grained ones. The number of time cost is in seconds. We provide detailed hardware and software configurations for experiments on time cost in Appendix \ref{appen:hardware_and_software_configurations}.}
\label{table:nq_tqa_retrieval_performance}
\end{table*}
\section{Experiment}
In this section, we conduct extensive experiments on two QA benchmark datasets to answer the following Research Questions (\textbf{RQs}): \textbf{RQ1:} How does progressive retrieval perform \textit{w.r.t.} Answer Recall when compared to flat one? \textbf{RQ2:} What are the benefits of performing \textsc{FunnelRAG} in Question Answering? \textbf{RQ3:} How do different settings influence the effectiveness of progressive retrieval? \textbf{RQ4:} How does the retrieval performance of \textsc{FunnelRAG} vary with different attention aggregation schemes? (Appendix \ref{appen:study_of_attention_aggregation_schema}) \textbf{RQ5:} Does ablating query tokens benefit the retrieval performance when selecting representative tokens? (Appendix \ref{appen:ablation_of_query_tokens}) \textbf{RQ6:} Does \textsc{FunnelRAG} contributes to a higher contextual integrity? (Appendix \ref{appen:contextual_integrity_analysis})
\subsection{Experimental Settings}
\smallsection{Datasets} We experiment on two QA datasets: \textit{i.e.,} Natural Question (NQ) \cite{DBLP:journals/tacl/KwiatkowskiPRCP19} and Trivia QA (TQA) \cite{DBLP:conf/acl/JoshiCWZ17}. Appendix \ref{appen:datasets} provides the statistics of the datasets. 
\smallsection{Metrics} Following \cite{DBLP:conf/nips/LewisPPPKGKLYR020} and \cite{DBLP:journals/corr/abs-2406-15319}, we employ Answer Recall (AR) and Exact Match (EM) to measure the performance of retrieval and generation, respectively. Besides, we use time cost as the metric of retrieval efficiency.
\smallsection{Retrievers} In progressive retrieval, we adopt BM25 \cite{DBLP:journals/ftir/RobertsonZ09} for retrieval, employ bge-reranker-v2-m3 \cite{DBLP:journals/corr/abs-2402-03216} as our pre-ranker, and leverage FiD \cite{DBLP:conf/eacl/IzacardG21} to perform post-ranking. As for flat retrieval, we use bge-large-en-v1.5 \cite{DBLP:journals/corr/abs-2309-07597}, the SOTA embedding model, to retrieve passages, and use bge-reranker-v2-m3\, to\, rerank. 
\smallsection{Generators} We choose two representative open-source LLMs: Llama3-8B-Instruct \cite{DBLP:journals/corr/abs-2407-21783} and Qwen2-7B-Instruct \cite{qwen2}, for our QA evaluation\footnote{We use Llama3-8B and Qwen2-7B to represent Llama3-8B-Instruct and Qwen2-7B-Instruct, respectively, for brevity.}. More details related to experimental settings can be found in Appendix \ref{appen:more_implementation_details}.
\subsection{Retrieval Performance (RQ1)}
\label{sec:retrieval_perf}
Table \ref{table:nq_tqa_retrieval_performance}, \ref{table:nq_tqa_retrieval_performance_top1}, \ref{table:nq_tqa_retrieval_performance_top2}, and \ref{table:nq_tqa_retrieval_performance_top3} shows the retrieval results on NQ and TQA datasets. 
As BM25's retrieval speed is breakneck, we mark its time cost as 0.00. From the results, we mainly have the following observations: 
\textbf{(1)} By adjusting the collaboration between large-to-small quantities, coarse-to-fine granularity, and simple-to-complex retrievers, progressive retrieval can find a gain-cost balance point.  
Specifically, the time cost of progressive retrieval is considerably lower than that of the flat one (reduced by about 40\%), while the AR is comparable to (or even better than) the flat one. 
The main reason for this results is that progressive retrieval combines the advantage of different retrievers but circumvents their disadvantage\footnote{Progressive retrieval combines simple retrievers' fast speed with complex ones' high precision, while it avoids simple retrievers' low precision and\, complex\, ones' slow\, speed.}.
\textbf{(2)} By retrieving in a funnel manner (\textit{e.g.,} 600K$\xrightarrow[]{\text{Retrieval}}$80c$\xrightarrow[]{\text{Pre-ranking}}$8d$\xrightarrow[]{\text{Post-ranking}}$4p),
progressive retrieval enables load balancing. It assigns simple but computationally demanding tasks to low-capacity retrievers, \textit{e.g.,} retrieving 80 clustered documents from 600K candidates. 
With almost no loss of golden units, the number of candidates dropped 7500x, largely increasing the signal-to-noise ratio. 
On the other hand, it assigns hard but computationally undemanding tasks to high-capacity retrievers, \textit{e.g.,} retrieving 4p passages from 8d document candidates. 
We provide more retrieval performance comparisons in Appendix \ref{appen:retrieval_performance_comparison}.
\begin{table}[!t]
\centering
\scriptsize
\tabcolsep=0.125cm
\renewcommand\arraystretch{1.2}
\begin{tabular}{ccclcl}
\toprule
    \multirow{2}{*}{\textbf{Generators}} &
    \multirow{2}{*}{\textbf{@K}} &
    \multicolumn{2}{|c}{\multirow{1}{*}{\bf\centering NQ}}
    &
    \multicolumn{2}{|c}{\multirow{1}{*}{\bf\centering TQA}}
    \\ 
    
    \cline{3-6}
    &  &  \multicolumn{1}{|c}{\centering Flat} & \multicolumn{1}{c}{Progressive} & \multicolumn{1}{|c}{\centering Flat} & \multicolumn{1}{c}{Progressive} \\ 
    \hline
    \hline
    \multirow{4}{*} {\centering {Llama3-8B}} & @1 & \multicolumn{1}{|c}{\centering 43.88} & \multicolumn{1}{l}{{\bf 51.27} (16.84\%)}  & \multicolumn{1}{|c}{\centering 67.54} & \multicolumn{1}{l}{\centering {\bf 70.77} (4.78\%)} \\
    & @2  & \multicolumn{1}{|c}{\centering 46.18} & \multicolumn{1}{l}{ {\bf  51.50} (11.52\%)}  & \multicolumn{1}{|c}{\centering 66.84} & \multicolumn{1}{l}{\centering {\bf 67.66} (1.23\%)}\\
    & @3  & \multicolumn{1}{|c}{\centering \bf  49.31 } & \multicolumn{1}{l}{ 49.06 (-0.51\%)}  & \multicolumn{1}{|c}{\centering 66.10} & \multicolumn{1}{l}{\centering {\bf 67.89} (2.71\%)}\\
    & @4  & \multicolumn{1}{|c}{\centering \bf  51.44} & \multicolumn{1}{l}{ 48.86 (-5.02\%)}  & \multicolumn{1}{|c}{\centering 67.10} & \multicolumn{1}{l}{\centering {\bf 68.47} (2.04\%)}\\
    \hline
    \multirow{4}{*} {\centering {Qwen2-7B}} & @1 & \multicolumn{1}{|c}{\centering 48.86} & \multicolumn{1}{l}{ {\bf  53.43} (9.35\%)}  & \multicolumn{1}{|c}{\centering 72.01} & \multicolumn{1}{l}{\centering {\bf 72.39} (0.53\%)} \\
    & @2 & \multicolumn{1}{|c}{\centering 53.32} & \multicolumn{1}{l}{ {\bf  54.27} (1.78\%)}  & \multicolumn{1}{|c}{\centering {\bf 73.69} } & \multicolumn{1}{l}{\centering 73.17 (-0.71\%)}\\
    & @3 & \multicolumn{1}{|c}{\centering 54.63} & \multicolumn{1}{l}{ {\bf  55.07} (0.81\%)}  & \multicolumn{1}{|c}{\centering \bf 74.60} & \multicolumn{1}{l}{\centering 73.73 (-1.17\%)}\\
    & @4 & \multicolumn{1}{|c}{\centering 55.46} & \multicolumn{1}{l}{ {\bf  55.48} (0.04\%)}  & \multicolumn{1}{|c}{\centering \bf 74.97} & \multicolumn{1}{l}{\centering 74.37 (-0.80\%)}\\
    \hline
    \hline
     \multicolumn{2}{c}{\multirow{1}{*}[-0.3ex]{Average Performance}} & \multicolumn{1}{|c}{\multirow{1}{*}[-0.3ex]{50.39}} & \multicolumn{1}{l}{\multirow{1}{*}[-0.3ex]{{\bf 52.37} (3.93\%)}}  & \multicolumn{1}{|c}{\multirow{1}{*}[-0.3ex]{70.36}} & \multicolumn{1}{l}{\multirow{1}{*}[-0.3ex]{{\bf 71.06} (0.99\%)}}\\
\bottomrule
\end{tabular}
\caption{Generation performance in terms of different retrieval paradigms. The \textbf{bold} indicates the best results.}
\label{table:nq_tqa_generation_performance}
\end{table}
\begin{figure*}[t]
    \centering  
     \subfigure[AR \textit{w.r.t.} different granularity.]{
        \includegraphics[width=.315\linewidth]{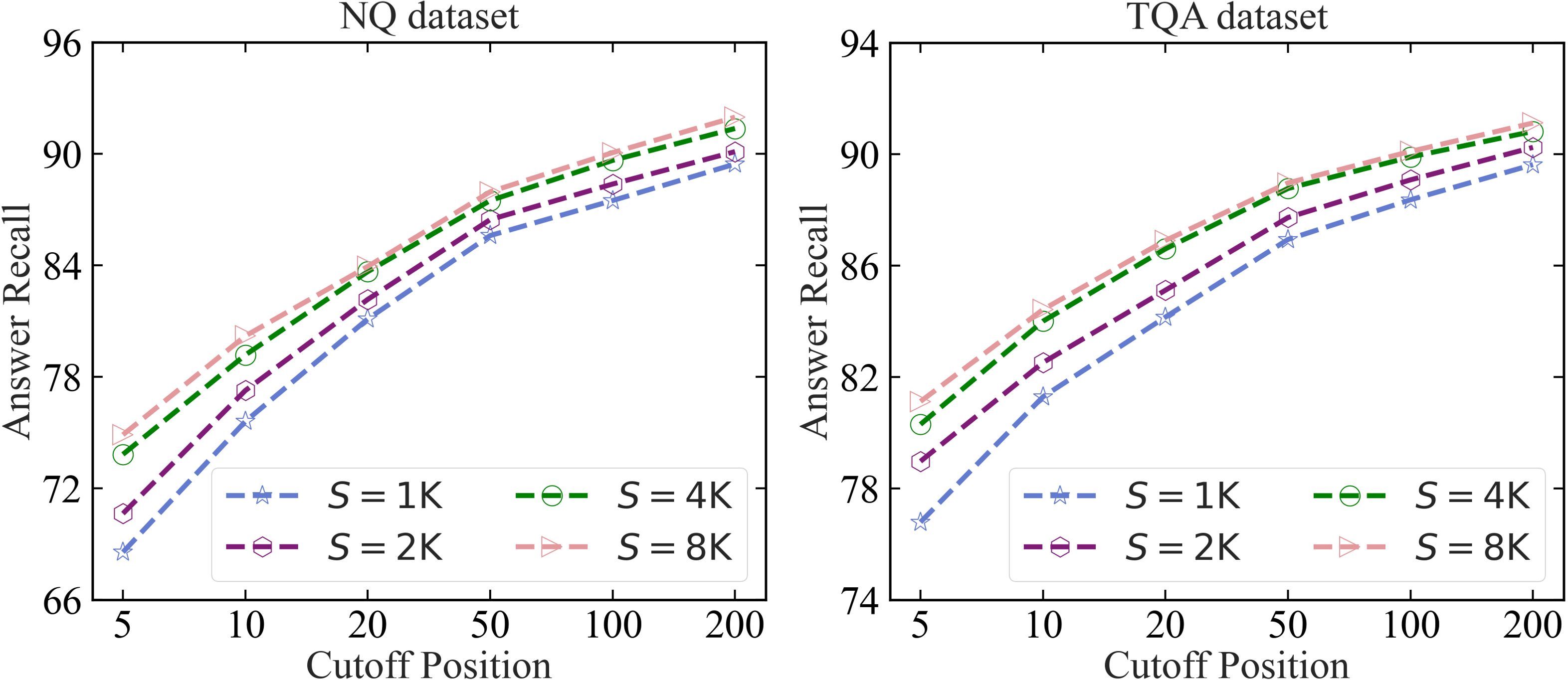}\label{fig:study_of_retrieval}}
    \subfigure[Impact of \#rep tokens (in Top-4).]{
        \includegraphics[width=.315\linewidth]{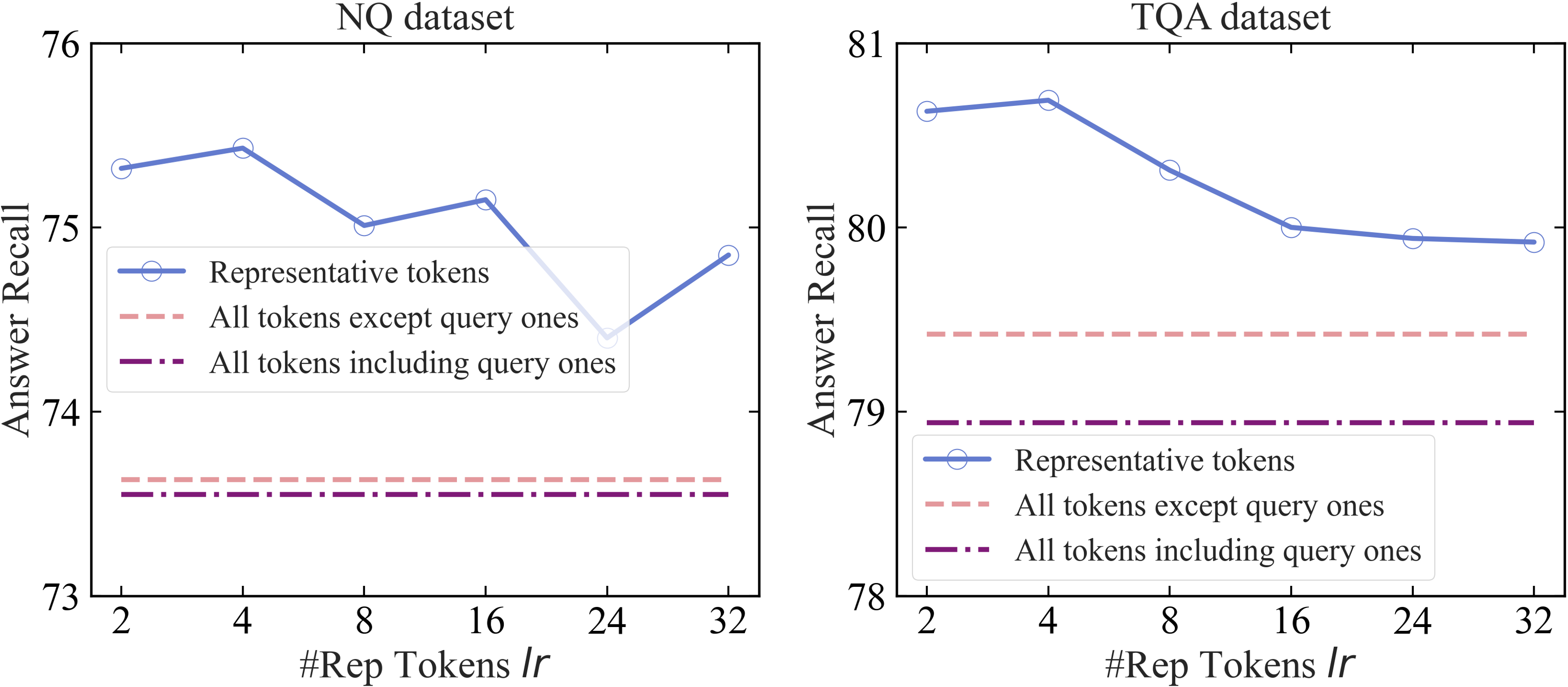}\label{fig:study_of_representative}}
     \subfigure[Effect of local-to-global distillation.]{
        \includegraphics[width=.3275\linewidth]{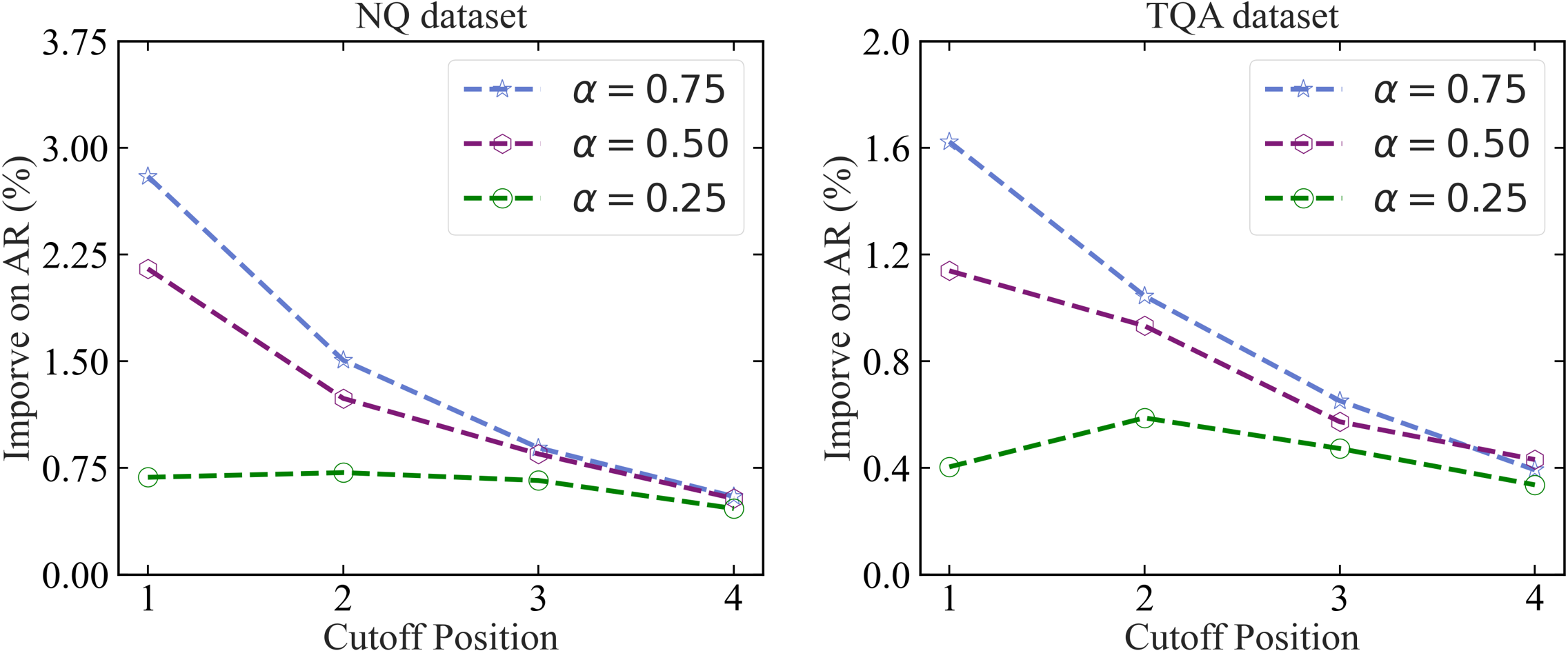}\label{fig:study_of_distill}}
    \caption{Model performance \textit{w.r.t.} (a) different granularity of clustered documents, (b) different number of representative tokens, and (c) L2G distillation. \#Rep tokens is the abbr of ``the number\, of representative tokens''.}
    \label{fig:study_of_funnelrag}
\end{figure*}
\begin{figure*}[t]
    \centering  
     \subfigure[Impact of \#rep tokens (in Top-1).]{
        \includegraphics[width=.315\linewidth]{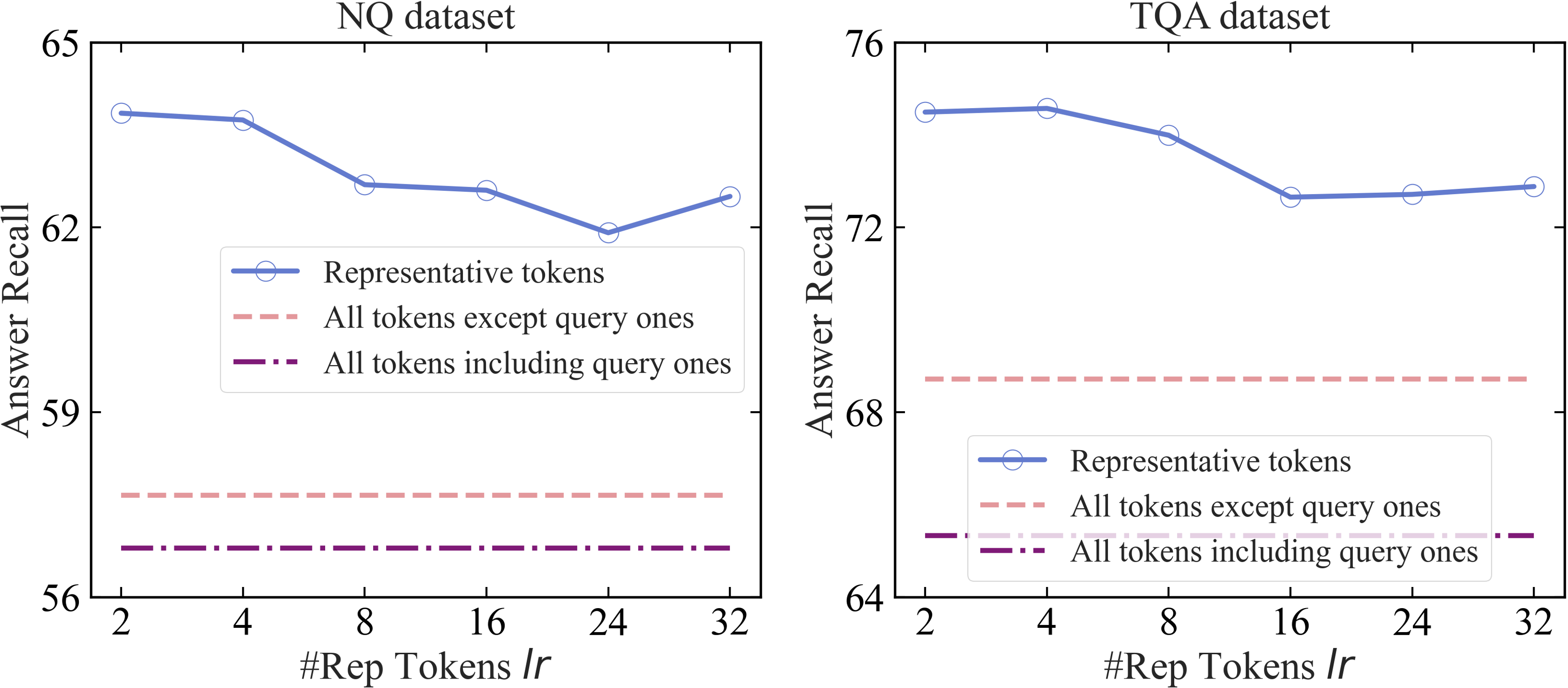}\label{fig:study_of_representative_t1}}
    \subfigure[Impact of \#rep tokens (in Top-2).]{
        \includegraphics[width=.315\linewidth]{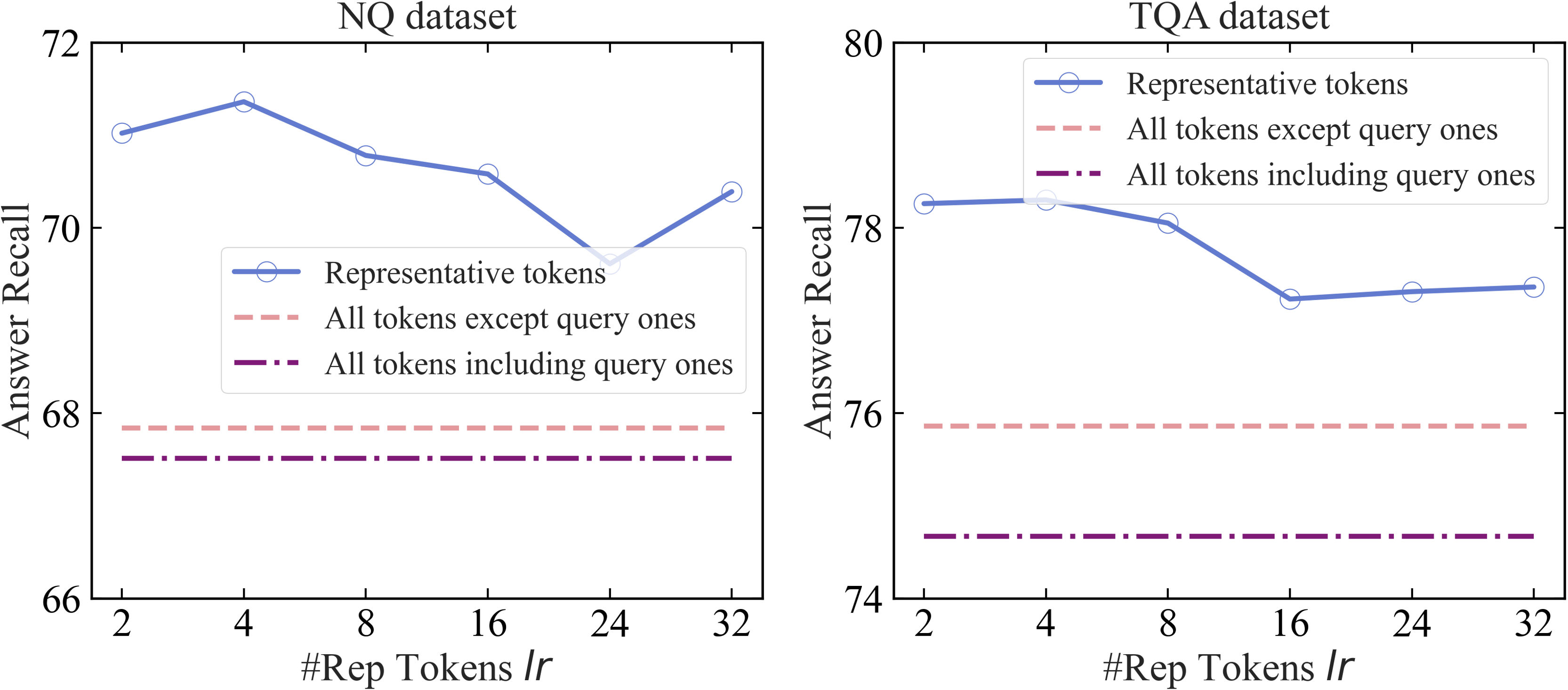}\label{fig:study_of_representative_t2}}
     \subfigure[Impact of \#rep tokens (in Top-3).]{
        \includegraphics[width=.315\linewidth]{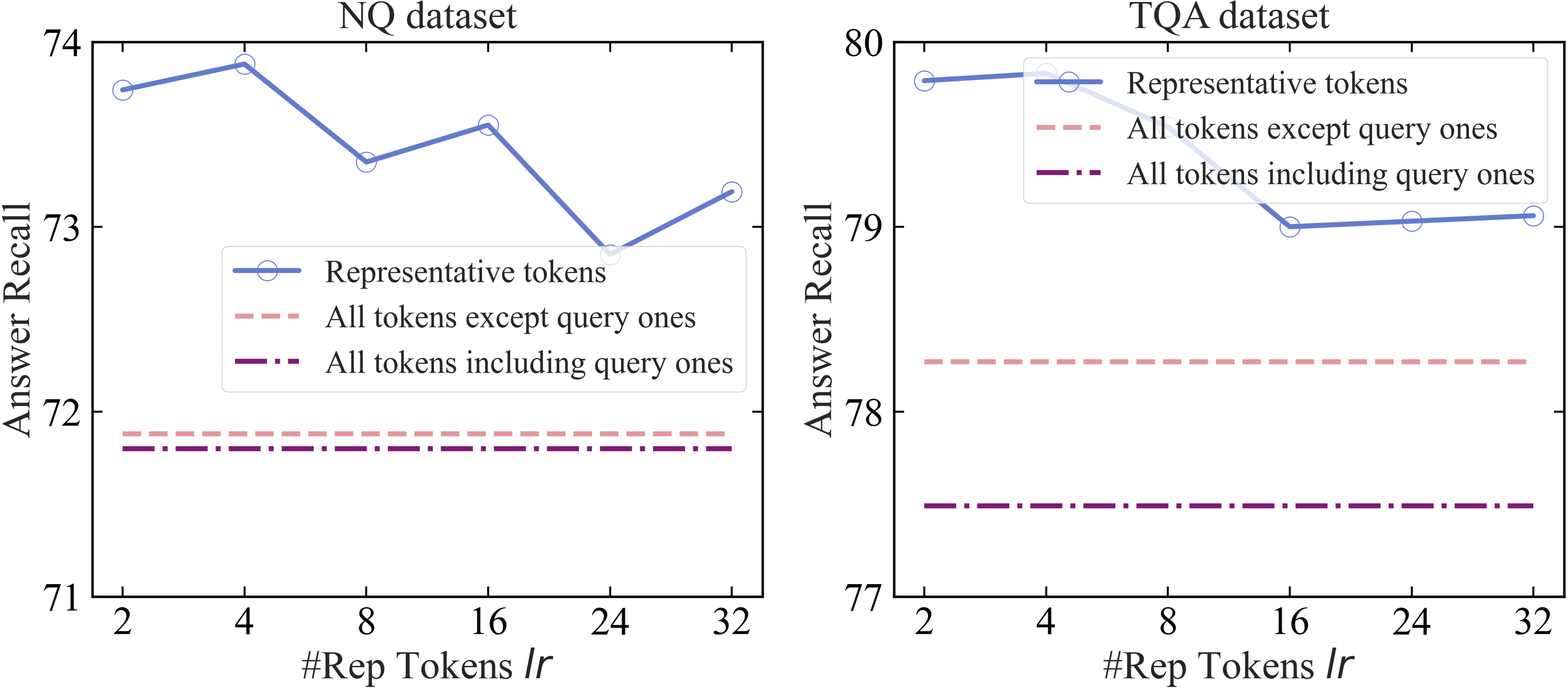}\label{fig:study_of_representative_t3}}
    \caption{AR comparison (in Top-1, Top-2, and Top-3) of \textsc{FunnelRAG} \textit{w.r.t.} different numbers of rep tokens.}
    \label{fig:rep_token_study}
\end{figure*}
\subsection{Generation Performance (RQ2)}
\label{sec:gen_performance}
Table \ref{table:nq_tqa_generation_performance} shows the generation performance \textit{w.r.t.} flat and progressive retrieval. 
From the results, we observe performance with progressive retrieval outperforms that with flat one in 13 out of 18 cases, indicating the superiority of supplementing LLMs with \textsc{FunnelRAG}. Specifically, when the cutoff position is small, progressive retrieval considerably outperforms flat one, such as the improvement of 16.87\% and 9.35\% on the NQ dataset in terms of @1 with Llama3-8B and Qwen2-7B, respectively. When the cutoff position is high, progressive retrieval still outperforms flat one in half of cases, even if the AR of progressive one is slightly lower than flat one. 
The reason is that passages retrieved by \textsc{FunnelRAG} have higher contextual integrity, refer to Appendix \ref{appen:contextual_integrity_analysis} for more details. Besides, we find Llama3-8B's performance may deteriorate when provided with more passages, whereas Qwen2-7B does not. The main reason may be the difference in their ability to handle long contexts.
\subsection{Study of \textsc{FunnelRAG} (RQ3)}
\label{sec:study_of_funnelrag}
In this section, we move on to studying different settings in the proposed \textsc{FunnelRAG} framework: 
{\textbf{(1) AR \textit{w.r.t.} different granularity:}} As shown in Figure \ref{fig:study_of_retrieval}, we experiment to evaluate the answer recall \textit{w.r.t.} different granularity of clustered documents, where we tune the max cluster size $S$ within the range of \{1K, 2K, 4K, 8K\}. It can be observed that setting $S$ as 8K does not bring much performance benefits compared to 4K. In contrast, setting $S$ as a small value (\textit{e.g.,} 1K) considerably degrades the performance. In view of this, we set the max cluster\, size $S$ as 4K\, tokens\, in this work.
{\textbf{(2) Impact of representative tokens:}} As shown in Figure \ref{fig:study_of_representative} and \ref{fig:rep_token_study}, we experiment to investigate \#rep tokens ${lr}$, where we search ${lr}$ in the range of \{2, 4, 8, 16, 24, 32\}.  We also aggregate FiD cross-attention scores across all tokens with/without query ones. We observe that the performance of aggregating rep tokens significantly outperforms that of aggregating all ones. On the other hand, it gets a peak value when selecting a few rep tokens, such as 4 in most cases. The results fully validate the necessity of selecting rep tokens.
More related details can be seen in Appendix \ref{appen:ablation_of_query_tokens}.
{\textbf{(3) Effect of local-to-global distillation:}} As shown in Figure \ref{fig:study_of_distill}, we experiment to explore the effect of local-to-global distillation, we tune $\alpha$ in the range of \{0.25, 0.5, 0.75\}. From the results, we observe that a higher value $\alpha$ (\textit{e.g.,} 0.75) usually leads to better performance and relatively high improvement on the top-ranked position (\textit{e.g.,} Top-1). The improvement is relatively trivial when $\alpha$ is small (\textit{e.g.,} 0.25). When the cutoff position is large (\textit{e.g.,} Top-4), different values of $\alpha$ bring similar performance improvements. In conclusion, we suggest tuning $\alpha$ within the range of [0.5, 1.0).

\section{Related Works}
\smallsection{Information Retrieval (IR)} IR models contain two categories: (1) Sparse retrieval, which computes relevance scores via lexical similarity \cite{DBLP:journals/ipm/SaltonB88}; (2) Dense retrieval, which captures semantic similarity within dense space \cite{DBLP:journals/tmlr/IzacardCHRBJG22,hu2025kalmembeddingsuperiortrainingdata}. To further improve retrieval performance, the re-ranking module is proposed to accurately re-estimate relevance scores using an enhanced model \cite{DBLP:conf/sigir/Zhuang0J0MLNWB23,DBLP:conf/emnlp/NogueiraJPL20,DBLP:journals/corr/abs-2407-03627}. 
Recently, LLMs have shown remarkable capability in re-ranking without further fine-tuning \cite{DBLP:journals/tmlr/LiangBLTSYZNWKN23,DBLP:conf/emnlp/0001YMWRCYR23,DBLP:journals/corr/abs-2309-15088}.
Despite effectiveness, we argue that they have not fully disclosed coarse-to-fine granularity evolution.
\smallsection{Retrieval-Augmented Generation} Retrieving knowledge from external sources to supplement LLMs' generation has been proven effective \cite{DBLP:conf/nips/LewisPPPKGKLYR020,DBLP:journals/corr/abs-2002-08909,DBLP:journals/tmlr/MialonDLNPRRSDC23,DBLP:journals/jmlr/IzacardLLHPSDJRG23,DBLP:conf/emnlp/ZhaoLZHCHZ24} in knowledge-intensive tasks \cite{DBLP:conf/naacl/PetroniPFLYCTJK21}. Specifically, RAG incorporates retrieval modules to provide external non-parametric knowledge into generation modules to remedy their incomplete, incorrect, or outdated internal parametric knowledge \cite{DBLP:conf/emnlp/KarpukhinOMLWEC20,DBLP:journals/corr/abs-1911-03868}.
Previous works jointly train the retrieval and generation modules \cite{DBLP:conf/icml/BorgeaudMHCRM0L22,DBLP:conf/nips/LewisPPPKGKLYR020,DBLP:journals/corr/abs-2002-08909}. With the rise of LLMs \cite{DBLP:journals/corr/abs-2303-08774}, most works directly serve them as generation modules without tuning due to their strong emergent abilities \cite{DBLP:journals/corr/abs-2402-13547,DBLP:conf/naacl/JeongBCHP24,DBLP:journals/corr/abs-2405-13792}. 
Inspired by RAG's strong practicality, a wide range of domains have developed specific RAG frameworks to perform their tasks,  including computer vision \cite{DBLP:journals/tmm/RaoDQFLST23,DBLP:journals/corr/abs-2405-10311,DBLP:journals/corr/abs-2411-07688,DBLP:journals/corr/abs-2411-16044}, knowledge graph \cite{DBLP:conf/acl/Yu0F0WXRY022,DBLP:journals/corr/abs-2404-16130}, and speech \cite{DBLP:journals/corr/abs-2406-03714,DBLP:conf/icassp/WangSS00YS24}. Albeit studied for ages, the balance between the effectiveness and efficiency of\, different\, retrievers\, is\, less\, explored.
\section{Conclusion}
In this paper, we propose a coarse-to-fine progressive retrieval paradigm for RAG, \textsc{FunnelRAG}, to enable load balancing and improve retrieval performance. The \textsc{FunnelRAG} framework efficiently retrieves useful passages by seamlessly collaborating different granularity (clustered document $\rightarrow$ passage), quantity (21M $\rightarrow$ 8d), and capacity (BM25 $\rightarrow$ FiD). 
Applying our framework reduces time cost by nearly 40\%, while the retrieval accuracy is comparable. In conclusion, our work sheds light on collaboration between complex and simple retrieval modules, and extensive experiments fully demonstrate\, its feasibility\, as\, well\, as\, usefulness.
\section*{Limitations}
Despite our innovations and improvements, we must acknowledge certain limitations in our work: 
\begin{itemize}[leftmargin=*]
    \item \textbf{Heuristic Metric.} The answer recall (AR), which is the main retrieval metric adopted in the experiments, might overestimate the usefulness of the retrieved information because it mechanically measures whether the retrieved information contains the answer string, even if the retrieved information does not convey accurate semantics. Evaluating metrics for RAG remains an \quad \quad open field that\, still\, needs\, further\, investigation.

    \item \textbf{Hand-crafted Labor.} Although our method is able to balance effectiveness and efficiency well, some hand-crafted labors are required to fulfill this goal. Specifically, to enable load balancing and improve retrieval accuracy, some hyper-parameters needed to be tuned, such as the max cluster size $S$, and the data flow between each stage required to carefully collaborate, such as the quantity of the retrieved units\, in each\, stage.

    \item \textbf{Compatibility Issue.} It is worth mentioning that \textsc{FunnelRAG} is a model-agnostic framework and can flexibly adapt to different retrievers. Therefore, the performance of \textsc{FunnelRAG} may be influenced by the characteristics of the retrievers. On the other hand, users can customize the retrievers to accommodate their specific scenarios. This work experiment on recently released retrievers and the results fully demonstrate\, the\, superiority\, of\, \textsc{FunnelRAG}.
\end{itemize}

\section*{Acknowledgments}
This work is jointly supported by grants: National Natural Science Foundation of China (No. 62376067), and Guangdong Basic and Applied Basic Research Foundation (2023A1515110078). We sincerely thank all anonymous reviewers for their detailed and careful reviews and valuable suggestions, which have significantly improved our work.

\bibliography{custom}

\clearpage

\appendix

\section{More Details Related to Methods}
\label{appen:more_details}
\subsection{Cross-Attention}
\label{appen:cross_attention}
Our model is built upon FiD \cite{DBLP:conf/eacl/IzacardG21}, whose architecture is a sequence-to-sequence model that consists of an encoder and a decoder. The encoder encodes each query-passage pair separately. The output representations of the encoder are concatenated (Equ (\ref{eq:fid})) along the sequence dimension to form a global representation $\mathbf{H} = \|_{m=1}^{M}\mathbf{H}_{q, p_m^f}$. 
Then, the decoder autoregressively attends to this representation, which alternates self-attention, cross-attention, and feed-forward modules. In the decoder, only the cross-attention module explicitly takes the global representation $\mathbf{H}$ of the encoder as the input. Let $\textbf{X}$ denotes the output of the previous self-attention layer of the decoder, the cross-attention operation consists of the following steps. First, three linear transformations are applied to produce queries $\mathbf{Q}$, keys $\mathbf{K}$, and values $\textbf{V}$, which can be formulated as:
\begin{equation}
    \mathbf{Q} = \mathbf{X}\mathbf{W}^Q, \mathbf{K} = \mathbf{H}\mathbf{W}^K, \mathbf{V} = \mathbf{H}\mathbf{W}^V.
\end{equation}
Then, the cross-attention scores are computed using the dot product between queries $\mathbf{Q}$ and keys $\mathbf{K}$. Taking the query at position $i$, $\mathbf{Q}_i$, and the key at position $j$, $\mathbf{K}_j$ for example, the cross-attention score and its normalized one can be\, computed\, as:
\begin{equation}
    \alpha_{i,j} = \mathbf{Q}_i \mathbf{K}_j,\quad\, \Tilde{\alpha}_{i,j} = \frac{\exp(\alpha_{i,j})}{\sum_k \exp(\alpha_{i,k})}.
\end{equation}
A new representation is derived from a sum of the values, weighted by the normalized cross-attention scores, going through a final linear transformation:
\begin{equation}
    \mathbf{O}_i = \mathbf{W}^O \sum_j \Tilde{\alpha}_{i,j} \mathbf{V}_{j}.
\end{equation}
We omit the layer and head indices for brevity in the above formula, refer to \cite{DBLP:conf/nips/VaswaniSPUJGKP17} for more details on the structure\, of\, Transformers.
\subsection{Representative Token Selection}
\label{appen:representative_token_selection}
Enlightened by the token redundancy in hidden states \cite{DBLP:conf/nips/DaiLY020,goyal2020power}, we select several representative tokens per layer and head, to estimate the relative importance of each passage more accurately. Specifically, we select ${lr}$ tokens that have the highest cross-attention scores from the passage to represent it. 
For the $i$-th layer and $j$-th head, we select the representative tokens according to their cross-attention scores and obtain an index list $\mathbf{r}$ that contains the index of ${lr}$ representative tokens. Then, we can average FiD cross-attention scores over representative tokens by looking\, up\, the\, index\, list\, $\mathbf{r}$, referring\, to\, Equ (\ref{equ:average_fid_score}).

\subsection{Training Procedure of FiD}
\label{app:fid}
Following \cite{DBLP:conf/eacl/IzacardG21}, we train FiD to generate the answer to the given query with the retrieved passages. Specifically, we adopt T5-large as the backbone of FiD, a generative encoder-decoder pretrained model. Then, each retrieved passage is concatenated with the query and processed separately by the T5 encoder. After that, the T5 decoder attends to the concatenation of the encoded representations of all retrieved passages. Finally, we train FiD to generate the answer via language modeling objective \cite{Radford2018ImprovingLU}.

\section{More Details Related to Experiments}
\label{appen:more_implementation_details}
\begin{table}[t]
  \centering
  \footnotesize
  \tabcolsep=0.185cm
  \renewcommand\arraystretch{1.25}
  \begin{tabular}{lccc}
    \toprule
    \multicolumn{1}{l}{\textbf{Dataset}} & 
    \multicolumn{1}{c}{\textbf{\#Train}} & \multicolumn{1}{c}{\textbf{\#Dev}} & \multicolumn{1}{c}{\textbf{\#Test}}\\
    \midrule
    NQ \cite{DBLP:journals/tacl/KwiatkowskiPRCP19}   & 79.1k & 8.7k & 3.6k \\
    TQA \cite{DBLP:conf/acl/JoshiCWZ17}  & 78.7k & 8.8k & 11.3k \\
    \bottomrule
  \end{tabular}
  \caption{\centering{Statistics of the datasets.}}
  \label{tab:dataset}
\end{table}
\begin{table*}[t]
    \renewcommand{\arraystretch}{1.2} 
    \begin{tcolorbox}
            \textbf{Retrieval-Augmented Question Answering Prompt} 
            \tcblower
            \textcolor{black}{\textbf{[Instruction]}}\\
            Given the [`context', `question'], answer the question based on the context.\\
            \textcolor{black}{\textbf{[Completion]}}\\
            {[context]:}  Nobel Prize in Physics The Nobel Prize in Physics () is a yearly award given by the Royal Swedish Academy of Sciences ...... \\
            {[question]:} Who got the first nobel prize in physics? \\
            {[answer]:} 
    \end{tcolorbox}
    \caption{The prompt for retrieval-augmented QA. For intuition, we select a real example from the NQ dataset.}
    \label{tab:gen_prompt}
\end{table*}
\subsection{Datasets}
\label{appen:datasets}
We experiment on NQ \cite{DBLP:journals/tacl/KwiatkowskiPRCP19} and TQA \cite{DBLP:conf/acl/JoshiCWZ17} datasets. Following \cite{DBLP:conf/acl/LeeCT19}, we discard answers with more than 5 tokens, as answers with many tokens often resemble extractive snippets instead of canonical ones. Table \ref{tab:dataset} shows the\, statistics\, of\, the\, datasets.
\subsection{LLMs for QA}
\label{appen:llms_for_qa}
\begin{itemize}[leftmargin=*]
    \item \textbf{Llama3-8B-Instruct}\footnote{\url{https://github.com/meta-llama/llama3}} is the developed version of Llama2 \cite{DBLP:journals/corr/abs-2307-09288}. Compared to Llama2, its training data volume has increased by seven times, showing higher reasoning and multilingual capabilities. For question answering, we use the Llama3-8B-Instruct, which is specifically optimized\, for\, the\, instruction-following\, ability.
    \item \textbf{Qwen2-7B-Instruct}\footnote{\url{https://github.com/QwenLM/Qwen2}} is the developed version of Qwen \cite{DBLP:journals/corr/abs-2309-16609}. Except for Chinese and English, its training data has also added high-quality data related to 27 languages and it achieves significant improvement in code and math capabilities. We also use the instruction-following version for the same reason aforesaid.
\end{itemize}

\subsection{Answer Recall}
\label{app:ar}
We measure retrieval performance with Answer Recall (AR) which is the recall of the answer string in all retrieved passages. It can be defined as follows:
\begin{equation}
    \mathrm{AR}@\mathrm{K} = \frac{1}{|\mathcal{Q}|}\sum_{q \in \mathcal{Q}} \mathbb{I}(a \in p_1\oplus p_2, ..., \oplus p_\mathrm{K}),
\end{equation}
where $@\mathrm{K}$ means retrieving Top-$\mathrm{K}$ relevant passages for each query, $\mathcal{Q}$ denotes the whole query set, $\mathbb{I}(\cdot)$ is an indicator function evaluating to 1 iff the answer string $a$ is included in the retrieved passages, $\oplus$ denotes the concatenation operation.
\subsection{Prompts}
\label{appen:prompts}
We provide the prompt that is used for retrieval-augmented question answering\, (\S \ref{sec:gen_performance}) in\, Table \ref{tab:gen_prompt}.
\subsection{Hardware and Software Configurations}
\label{appen:hardware_and_software_configurations}
\smallsection{Hardware Configurations} The experiments are conducted on a Linux server equipped with an AMD EPYC 7742 64-Core Processor, 1000GB RAM, and\, 2 NVIDIA A100-SXM4-40GB\, GPUs.
\smallsection{Software Configurations} We run retrieval and generation experiments on the above hardware. We conduct experiments with the FlagEmbedding\footnote{\url{https://github.com/FlagOpen/FlagEmbedding}} and huggingface transformers toolkit \cite{DBLP:conf/emnlp/WolfDSCDMCRLFDS20}. 
For dense retrieval, we leverage the open-source retrieval toolkit Tevatron  \cite{DBLP:journals/corr/abs-2203-05765}. On the other hand, we adopt BM25S \cite{bm25s} for sparse retrieval. Besides, we adopt T5-large\footnote{\url{https://huggingface.co/google-t5/t5-large}} as the backbone of FiD (\S\ref{sec:post_ranking_stage}). The detailed setting of the software environment is provided in Table \ref{tab:software}.
\begin{table}[h]
    \footnotesize
    \renewcommand{\arraystretch}{1.25} 
    \tabcolsep=0.125cm
    \centering
    \begin{tabular}{cc}
    \toprule
    \textbf{Configuration} & \textbf{Value} \\ 
    \midrule
        Tevatron & V2 \\
        bm25s & 0.1.10 \\
        faiss-cpu & 1.8.0.post1 \\
        FlagEmbedding & 1.2.11 \\
        transformers & 4.44.2 \\
    \bottomrule
    \end{tabular}
    \caption{Detailed settings of the software environment.}
    \label{tab:software}
\end{table}
\begin{table*}[!t]
\centering
\footnotesize
\tabcolsep=0.175cm
\renewcommand\arraystretch{1.25}
\begin{tabular}{cccccc|ccccc}
\toprule
    \multirow{2}{*}{\textbf{Method}} &
    \multicolumn{5}{|c}{\textbf{NQ}} &
    \multicolumn{5}{|c}{\textbf{TQA}} 
    \\ 
    \cline{2-11}
    &  \multicolumn{1}{|c}{\centering @1} & \multicolumn{1}{c}{@2} &  \multicolumn{1}{c}{\centering @3} & \multicolumn{1}{c}{@4} & \multicolumn{1}{c}{Avg} &   \multicolumn{1}{|c}{\centering @1} & \multicolumn{1}{c}{@2} &  \multicolumn{1}{c}{\centering @3} & \multicolumn{1}{c}{@4} & \multicolumn{1}{c}{Avg}\\ 
    \hline
    \hline
    \multicolumn{1}{l}{(\romannumeral1) $\text{mean}_{i,j,k} \alpha_{0,i,j,\mathbf{r}[k]}$} & \multicolumn{1}{|c}{\centering \bf 63.74} & \multicolumn{1}{c}{\centering \bf 71.36} & \multicolumn{1}{c}{\centering \bf 73.88} & \multicolumn{1}{c}{\centering 75.43} &  \multicolumn{1}{c}{\centering \bf 71.10} & \multicolumn{1}{|c}{\centering 74.57} & \multicolumn{1}{c}{\centering \bf 78.31} & \multicolumn{1}{c}{\centering 79.83} & \multicolumn{1}{c}{\centering 80.69} &  \multicolumn{1}{c}{\centering 78.35} \\
    
    \multicolumn{1}{l}{(\romannumeral2) $\text{mean}_{j,k} \text{max}_{i} \alpha_{0,i,j,k}$} & \multicolumn{1}{|c}{\centering 57.06} & \multicolumn{1}{c}{\centering 68.01} & \multicolumn{1}{c}{\centering 71.91} & \multicolumn{1}{c}{\centering 73.85} &  \multicolumn{1}{c}{\centering 67.71} & \multicolumn{1}{|c}{\centering 66.98} & \multicolumn{1}{c}{\centering 75.44} & \multicolumn{1}{c}{\centering 78.02} & \multicolumn{1}{c}{\centering 79.56} &  \multicolumn{1}{c}{\centering 75.00}\\

    \multicolumn{1}{l}{(\romannumeral3) $\text{mean}_{i,k} \text{max}_{j} \alpha_{0,i,j,k}$} & \multicolumn{1}{|c}{\centering 57.65} & \multicolumn{1}{c}{\centering 68.14} & \multicolumn{1}{c}{\centering 71.86} & \multicolumn{1}{c}{\centering 73.77} &  \multicolumn{1}{c}{\centering 67.86} & \multicolumn{1}{|c}{\centering 68.32} & \multicolumn{1}{c}{\centering 75.70} & \multicolumn{1}{c}{\centering 78.12} & \multicolumn{1}{c}{\centering 79.47} &  \multicolumn{1}{c}{\centering 75.40}\\

    \multicolumn{1}{l}{(\romannumeral4) $\text{mean}_{i,j} \text{max}_{k} \alpha_{0,i,j,k}$} & \multicolumn{1}{|c}{\centering 63.32} & \multicolumn{1}{c}{\centering 70.69} & \multicolumn{1}{c}{\centering 73.52} & \multicolumn{1}{c}{\centering 75.12} &  \multicolumn{1}{c}{\centering 70.66} & \multicolumn{1}{|c}{\centering 74.10} & \multicolumn{1}{c}{\centering 78.11} & \multicolumn{1}{c}{\centering 79.70} & \multicolumn{1}{c}{\centering 80.59} &  \multicolumn{1}{c}{\centering 78.13}\\

    \multicolumn{1}{l}{(\romannumeral5) $\text{mean}_{-6\leq i \leq -1,j,k} \alpha_{0,i,j,\mathbf{r}[k]}$} & \multicolumn{1}{|c}{\centering 63.68} & \multicolumn{1}{c}{\centering 71.16} & \multicolumn{1}{c}{\centering 73.82} & \multicolumn{1}{c}{\centering \bf  75.51} &  \multicolumn{1}{c}{\centering 71.04} & \multicolumn{1}{|c}{\centering \bf 74.64} & \multicolumn{1}{c}{\centering 78.26} & \multicolumn{1}{c}{\centering \bf 79.93} & \multicolumn{1}{c}{\centering \bf 80.89} &  \multicolumn{1}{c}{\centering \bf 78.43}\\

\bottomrule
\end{tabular}
\caption{Performance comparison of attention aggregation schemes on NQ and TQA datasets, where the index $i$ corresponds to layers of the decoder, $j$ corresponds to heads, and $k$ corresponds to input tokens of a given passage.}
\label{table:agg_schema_fid}
\end{table*}
\section{More In-depth Studies}
\begin{table}[!t]
\centering
\footnotesize
\tabcolsep=0.15cm
\renewcommand\arraystretch{1.25}
\begin{tabular}{ccccc}
\toprule
    \multirow{1}{*}{\textbf{Datasets}} &
    \multirow{1}{*}{\textbf{@K}} &
    \multicolumn{1}{|c}{\textbf{w/ query}} &
    \multicolumn{1}{c}{\textbf{w/o query}} &
    \multicolumn{1}{|c}{\textbf{\%Improv.}} 
    \\
    \hline
    \hline
    \multirow{4}{*} {\centering {NQ}} & @1 & \multicolumn{1}{|c}{\centering 56.79 } & \multicolumn{1}{c}{\bf 57.65} & \multicolumn{1}{|c}{\centering 1.51\%}  \\
     & @2 & \multicolumn{1}{|c}{\centering 67.51} & \multicolumn{1}{c}{\bf 67.84} & \multicolumn{1}{|c}{\centering 0.49\%} \\
     & @3 & \multicolumn{1}{|c}{\centering 71.80} & \multicolumn{1}{c}{\bf 71.88} & \multicolumn{1}{|c}{\centering 0.11\%} \\
     & @4 & \multicolumn{1}{|c}{\centering 73.55} & \multicolumn{1}{c}{\bf 73.63} & \multicolumn{1}{|c}{\centering 0.11\%} \\
     \hline
     \multirow{4}{*} {\centering {TQA}} & @1 & \multicolumn{1}{|c}{\centering 65.33} & \multicolumn{1}{c}{\bf 68.71} & \multicolumn{1}{|c}{\centering 5.17\%}  \\
     & @2 & \multicolumn{1}{|c}{\centering 74.67} & \multicolumn{1}{c}{\bf 75.86} & \multicolumn{1}{|c}{\centering 1.59\%} \\
     & @3 & \multicolumn{1}{|c}{\centering 77.49} & \multicolumn{1}{c}{\bf 78.27} & \multicolumn{1}{|c}{\centering 1.01\%} \\
     & @4 & \multicolumn{1}{|c}{\centering 78.94} & \multicolumn{1}{c}{\bf 79.42} & \multicolumn{1}{|c}{\centering 0.61\%} \\
\bottomrule
\end{tabular}
\caption{AR comparison with and without query tokens.}
\label{table:ablation_query}
\end{table}
\begin{table}[t]
  \centering
  \footnotesize
  \tabcolsep=0.125cm
  \renewcommand\arraystretch{1.25}
  \begin{tabular}{ccc}
    \toprule
    \multicolumn{1}{c}{\textbf{Dataset}} & 
    \multicolumn{1}{c}{\textbf{Flat Retrieval}} & \multicolumn{1}{c}{\textbf{ Progressive Retrieval}}\\
    \midrule
    NQ    & 1.330 & \textbf{1.259} \\
    TQA   & 1.760 & \textbf{1.494} \\
    \bottomrule
  \end{tabular}
  \caption{{Contextual integrity measurement on passages. The lower the value, the better the\, contextual\, integrity.}}
  \label{tab:contextual_analysis}
\end{table}
\subsection{Study of Attention Aggregation Schema (RQ4)}
\label{appen:study_of_attention_aggregation_schema}
Table \ref{table:agg_schema_fid} shows the results with different aggregation schemes. Particularly, we consider \textbf{(1)} taking the average over all the layers, all the heads, and all the representative tokens; \textbf{(2)}  taking the max over the layers instead of the average; \textbf{(3)} taking the max over the heads instead of the average; \textbf{(4)} taking the max over input tokens instead of the average; \textbf{(5)}  taking the mean over the last 6 layers instead of all the layers. 
In all of the above variants, we do not consider query tokens when aggregating. We observe that: (1) Aggregation schemes with representative tokens selection (\textit{i.e.,} (\romannumeral1), (\romannumeral4), and (\romannumeral5)) outperform those without representative tokens selection by a large margin, which indicates the necessity of selecting representative tokens. (2) Comparing (\romannumeral1) and (\romannumeral4), we find that setting \#rep tokens to a small value (\textit{e.g.,} 1) will hurt the retrieval performance, which suggests setting \#rep tokens to moderate values (\textit{e.g.,} 4). (3) Comparing (\romannumeral1) and (\romannumeral5), only considering part of the decoder layers may bring a little performance degradation. The main reason is that certain clues captured by the other layers have been neglected. (4) Comparing (\romannumeral2) and (\romannumeral3), taking the max over the heads generally performs better than that over the layers. 
The main reason is taking the max over the heads may find the influential head that plays a key role in identifying clues \cite{DBLP:journals/corr/abs-2406-11497}, while taking the max over the layers will miss some key clues captured by the other layers. In conclusion, we suggest setting \#rep tokens to moderate values, using all the layers, and trying to identify the influential heads.
\subsection{Ablation of Query Tokens (RQ5)}
\label{appen:ablation_of_query_tokens}
In Table \ref{table:ablation_query}, we conduct experiments to verify the effectiveness of ablating query tokens when estimating the relative importance of each passage. Averaging the  FiD cross-attention score of all tokens in $q\oplus p^f$ mentions ``\textbf{w/ query}'' and averaging that of all tokens in $p^f$ mentions ``\textbf{w/o query}''. Here, we do not consider selecting representative tokens so that we can verify the effectiveness of ablating query tokens straightforwardly. The results show that ablating query tokens consistently improves answer recall across different cutoff positions. Specifically, We find that the marginal benefit appears as the cutoff position increases. 
For example, on the TQA dataset, \textbf{w/o query} achieves 5.17\% improvement over \textbf{w/ query} at the Top-1 position, while it is 0.61\% at the Top-4 position. It can be concluded that ablating query tokens can significantly improve answer recall in the top-ranked positions, however, the gain decreases as the number of passages retrieved increases. The main reason is that as the number of passages retrieved increases, it becomes more essential than ablating query tokens.
\subsection{Contextual Integrity Analysis (RQ6)}
\label{appen:contextual_integrity_analysis}
Generally, a higher contextual integrity is in favor of most downstream tasks. To measure contextual integrity, we compute information entropy \cite{DBLP:journals/sigmobile/Shannon01} on passages' source in terms of Top-4 passages for each query, where we use the title of passages as the identifier. Table \ref{tab:contextual_analysis} shows the measurement results \textit{w.r.t.} flat and progressive retrieval. 
From the results, we find that progressive retrieval achieves lower information entropy (\textit{i.e.,} higher contextual integrity) than flat one. Specifically, the improvement of progressive retrieval over flat one is 5.34\% and 15.11\% in terms of NQ and TQA datasets, respectively. The results fully verify that the proposed \textsc{FunnelRAG} can largely enhance the contextual integrity just as we discussed in \S\ref{sec:intro}.
\begin{table*}[t]
    \centering
    \small
    \renewcommand{\arraystretch}{1.4} 
    \begin{tabular}{p{0.96\linewidth}} 
        \toprule
        \textbf{Question:} Who won the fifth season of America's Got Talent? \\
        \textbf{Answer:} [``Soul singer Michael Grimm'', ``Michael Grimm''] \\
        \midrule 
        \textbf{(1) Retrieval Stage (clustered documents):} (..., Miss Kansas USA, America's Got Talent (season 9), Hurricane (Nick Fradiani album), ...); (..., Preacher Lawson, \textbf{\textcolor[RGB]{0,160,0}{America's Got Talent}}, Jodi Miller, ...); (..., Kite line, \textbf{\textcolor[RGB]{0,160,0}{Michael Grimm (album)}}, \textbf{\textcolor[RGB]{0,160,0}{America's Got Talent (season 5)}}, \textbf{\textcolor[RGB]{0,160,0}{Michael Grimm (musician)}}, Connor Doran, ...); (America's Got Talent (season 4), Kevin Skinner, The Spiritual Harmonizers, ...) ...... \\
        \midrule 
        \textbf{(2) Pre-ranking Stage (documents):} \textbf{\textcolor[RGB]{0,160,0}{America's Got Talent (season 5)}}; \textbf{\textcolor[RGB]{0,160,0}{Michael Grimm (musician)}}; America's Got Talent (season 12); America's Got Talent (season 13); \textbf{\textcolor[RGB]{0,160,0}{America's Got Talent}}; Got Talent; America's Got Talent (season 9) ...... \\
        \midrule 
        \textbf{(3) Post-ranking Stage (passages):} \textbf{\textcolor[RGB]{0,160,0}{America's Got Talent (season 5)}}; \textbf{\textcolor[RGB]{0,160,0}{Michael Grimm (musician)}}; America's Got Talent; \textbf{\textcolor[RGB]{0,160,0}{America's Got Talent (season 5)}}. \\
        \bottomrule
    \end{tabular}
    \caption{An example of \textsc{FunnelRAG}'s data flow from NQ dataset, where the retrieved information that contains the answer string is marked in \textbf{\textcolor[RGB]{0,160,0}{green}}. Note\, that\, we\, only\, present\, the\, title\, of\, the\, retrieved information for brevity.}
    \label{tab:case_study_prog}
\end{table*}
\subsection{Retrieval Performance Comparison}
\label{appen:retrieval_performance_comparison}
Table \ref{table:nq_tqa_retrieval_performance_top1}-\ref{table:nq_tqa_retrieval_performance_top3} shows the retrieval performance on NQ and TQA datasets in terms of the cutoff position of 1, 2, and 3. The results show similar trends as Table \ref{table:nq_tqa_retrieval_performance}. Specifically, progressive retrieval usually achieves better performance while using less time cost. 
For example, on the NQ dataset, progressive retrieval uses the time cost of 2.97s and obtains the \textbf{AR} of 63.73, in terms of Top-1, while flat one takes 5.25s and only achieves the \textbf{AR} of 56.09. The results verify the superiority of progressive retrieval in balancing effectiveness and efficiency.
\subsection{Case Study}
\label{appen:case_study}
Table \ref{tab:case_study_prog} shows a data-flow example of \textsc{FunnelRAG} from the NQ dataset. It consists of three stages: Retrieval, Pre-ranking, and Post-ranking. From the retrieval to the post-ranking stage, the granularity evolves from coarse-grained clustered documents to fine-grained passages. Besides, the number of candidates decreases step by step (\textit{e.g.,} 600K$\xrightarrow[]{\text{Retrieval}}$80c$\xrightarrow[]{\text{Pre-ranking}}$8d$\xrightarrow[]{\text{Post-ranking}}$4p). More importantly, we find that the proportion of candidates that contains the answer string gradually increases, which indicates that the signal-to-noise ratio is greatly improved through progressive retrieval.
Furthermore, with more advanced models, the pre-ranking and the post-ranking stage can perceive clues in the question, \textit{i.e.,} ``\textbf{the fifth season}''. Though the model becomes more complex, the number of candidates is limited, so the computational cost is small. As a result, ``\textbf{America’s Got Talent (season 5)}'' has been ranked first. The above analyses fully verify the effectiveness and reasonability of the proposed retrieval paradigm for RAG, \textit{i.e.,} \textsc{FunnelRAG}, whose main design ideas are large-to-small quantities, coarse-to-fine granularity, as well as simple-to-complex\, models.
\renewcommand{\algorithmicrequire}{\textbf{Input:}}  
\renewcommand{\algorithmicensure}{\textbf{Output:}}  
\begin{algorithm}[t]
\caption{Document Clustering Algorithm}
\label{alg:cluster_documents}
\begin{algorithmic}[1]
    \Require $\mathcal{D}$ (documents), $S$ (max cluster size), $\text{adj}[d]$ (related documents for each $d$)
    \Ensure $\mathcal{C}$ (set of clusters)
    \State Sort $\mathcal{D}$ by their local cluster coefficient;
    \State Initialize an empty set of clusters $\mathcal{C} \gets \emptyset$;
    \For{each document $d$ in $\mathcal{D}$}
        \State Add $\{d\}$ to $\mathcal{C}$;
    \EndFor
    \For{each document $d$ in $\mathcal{D}$}
        \State Remove $\{d\}$ from $\mathcal{C}$;
        \State $\mathcal{R}\gets\textsc{find\_related\_cluster}(d,\mathcal{C})$;
        \State Create a new cluster $c_{\text{new}} = \{d\}$;
        \State Sort $\mathcal{R}$ by their closeness centrality to $d$;
        \For{each cluster c in $\mathcal{R}$}
            \If{$|c_{\text{new}}|+|c|\leq S$}
                \State $c_{\text{new}} \gets c_{\text{new}} \cup c $;
                \State Remove $c$ from $\mathcal{C}$;
            \EndIf
        \EndFor
        \State Add $c_{\text{new}}$ to $\mathcal{C}$;
    \EndFor
    \State \Return $\mathcal{C}$;
    \State 
    \Function{\textsc{find\_related\_cluster}}{$d$, $\mathcal{C}$}
        \State Initialize the set of related clusters $\mathcal{R} \gets \emptyset$;
        \For{each related document $d^\prime$ in $\text{adj}[d]$}
            \For{each cluster $c$ in $\mathcal{C}$}
                \If{$d^\prime \in c$}
                    \State $\mathcal{R} \gets \mathcal{R} \cup \{c\}$;
                \EndIf
            \EndFor
        \EndFor
        \State \Return The set of related clusters $\mathcal{R}$;
    \EndFunction  
\end{algorithmic}
\end{algorithm}
\begin{table*}[!ht]
\centering
\footnotesize
\tabcolsep=0.125cm
\renewcommand\arraystretch{1.25}
\begin{tabular}{ccllllc}
\toprule
    \multirow{2}{*}[-0.5ex]{\textbf{Datasets}} &
    \multirow{2}{*}[-0.5ex]{\textbf{Retrieval Paradigm}} &
    \multicolumn{3}{c}{\textbf{Retrieval Stages}} &
    \multicolumn{1}{c}{\multirow{2}{*}[-0.4ex]{\textbf{\begin{tabular}[c]{@{}c@{}}Time Cost \\ (T)\end{tabular}}}} &
    \multirow{2}{*}[-0.4ex]{\textbf{\begin{tabular}[c]{@{}c@{}}Answer Recall \\ (AR)\end{tabular}}} \\ 
    \cmidrule(lr){3-5}
    &  &  \multicolumn{1}{c}{\centering Retrieval} & \multicolumn{1}{c}{Pre/Re-ranking} & \multicolumn{1}{c}{Post-ranking} &  &  \\ 
    \midrule

    & \multirow{2}{*}{Flat Retrieval}  &   21M $\rightarrow$ 1p  &  N/A  &  N/A   &  4.90 (4.90+N/A+N/A) & 51.16 \\
    & & 21M $\rightarrow$ 400p & 400p $\rightarrow$ 1p &  N/A &  5.25 (4.90+0.35+N/A) & 56.09 \\
    \rowcolor{gray!12}
    \cellcolor{white!20} &  & 600K $\rightarrow$ 1c & N/A  &  N/A  &  0.00 (0.00+N/A+N/A) &  52.74 \\
    \rowcolor{gray!12}
   \cellcolor{white!20} &  & 600K $\rightarrow$ 20c & 20c $\xrightarrow[]{\sim \text{180d}}$ 1d  &  N/A  &  0.49 (0.00+0.49+N/A) & 63.46   \\
    \rowcolor{gray!12}
    \cellcolor{white!20} \multirow{-5}{*}{\centering {NQ}} & \multirow{-3}{*}[0.5ex]{Progressive Retrieval} & 600K $\rightarrow$ 80c & 80c $\xrightarrow[]{\sim \text{740d}}$ 8d & 8d $\xrightarrow[]{\sim \text{50p}}$ 1p & 2.97 (0.00+2.20+0.77) & 63.74 \\
    \midrule
     & \multirow{2}{*}{Flat Retrieval}  &   21M $\rightarrow$ 1p  &  N/A  &  N/A   &  5.02 (5.02+N/A+N/A) & 60.29  \\
    &  & 21M $\rightarrow$ 400p  &  400p $\rightarrow$ 1p  &  N/A   &  5.41 (5.02+0.39+N/A) & 71.37 \\
    \rowcolor{gray!12}
    \cellcolor{white!20} &  & 600K $\rightarrow$ 1c & N/A  &  N/A  &  0.00 (0.00+N/A+N/A) &  66.48 \\
    \rowcolor{gray!12}
    \cellcolor{white!20} &  & 600K $\rightarrow$ 20c & 20c $\xrightarrow[]{\sim \text{200d}}$ 1d  &  N/A  &  0.60 (0.00+0.60+N/A) &  72.59 \\
    \rowcolor{gray!12}
    \cellcolor{white!20} \multirow{-5}{*} {\centering {TQA}} & \multirow{-3}{*} {\centering  Progressive Retrieval} & 600K $\rightarrow$ 80c & 80c $\xrightarrow[]{\sim \text{800d}}$ 12d & 12d $\xrightarrow[]{\sim \text{65p}}$ 1p & 3.47 (0.00+2.52+0.95) & 74.57 \\

\bottomrule
\end{tabular}
\caption{Retrieval performance comparison (in Top-1) \textit{w.r.t.} time cost and answer recall on NQ and TQA datasets.}
\label{table:nq_tqa_retrieval_performance_top1}
\end{table*}
\begin{table*}[!ht]
\centering
\footnotesize
\tabcolsep=0.125cm
\renewcommand\arraystretch{1.25}
\begin{tabular}{ccllllc}
\toprule
    \multirow{2}{*}[-0.5ex]{\textbf{Datasets}} &
    \multirow{2}{*}[-0.5ex]{\textbf{Retrieval Paradigm}} &
    \multicolumn{3}{c}{\textbf{Retrieval Stages}} &
    \multicolumn{1}{c}{\multirow{2}{*}[-0.4ex]{\textbf{\begin{tabular}[c]{@{}c@{}}Time Cost \\ (T)\end{tabular}}}} &
    \multirow{2}{*}[-0.4ex]{\textbf{\begin{tabular}[c]{@{}c@{}}Answer Recall \\ (AR)\end{tabular}}} \\ 
    \cmidrule(lr){3-5}
    &  &  \multicolumn{1}{c}{\centering Retrieval} & \multicolumn{1}{c}{Pre/Re-ranking} & \multicolumn{1}{c}{Post-ranking} &  &  \\ 
    \midrule

    & \multirow{2}{*}{Flat Retrieval}  &   21M $\rightarrow$ 2p  &  N/A  &  N/A   &  4.90 (4.90+N/A+N/A) & 64.21 \\
    & & 21M $\rightarrow$ 400p & 400p $\rightarrow$ 2p &  N/A &  5.25 (4.90+0.35+N/A) & 67.34 \\
    \rowcolor{gray!12}
    \cellcolor{white!20} &  & 600K $\rightarrow$ 2c & N/A  &  N/A  &  0.00 (0.00+N/A+N/A) &  63.57 \\
    \rowcolor{gray!12}
   \cellcolor{white!20} &  & 600K $\rightarrow$ 20c & 20c $\xrightarrow[]{\sim \text{180d}}$ 2d  &  N/A  &  0.49 (0.00+0.49+N/A) & 70.17   \\
    \rowcolor{gray!12}
    \cellcolor{white!20} \multirow{-5}{*}{\centering {NQ}} & \multirow{-3}{*}[0.5ex]{Progressive Retrieval} & 600K $\rightarrow$ 80c & 80c $\xrightarrow[]{\sim \text{740d}}$ 8d & 8d $\xrightarrow[]{\sim \text{50p}}$ 2p & 2.97 (0.00+2.20+0.77) & 71.36 \\
    \midrule
     & \multirow{2}{*}{Flat Retrieval}  &   21M $\rightarrow$ 2p  &  N/A  &  N/A   &  5.02 (5.02+N/A+N/A) & 68.98  \\
    &  & 21M $\rightarrow$ 400p  &  400p $\rightarrow$ 2p  &  N/A   &  5.41 (5.02+0.39+N/A) & 77.29 \\
    \rowcolor{gray!12}
    \cellcolor{white!20} &  & 600K $\rightarrow$ 2c & N/A  &  N/A  &  0.00 (0.00+N/A+N/A) &  73.58 \\
    \rowcolor{gray!12}
    \cellcolor{white!20} &  & 600K $\rightarrow$ 20c & 20c $\xrightarrow[]{\sim \text{200d}}$ 2d  &  N/A  &  0.60 (0.00+0.60+N/A) &  77.13 \\
    \rowcolor{gray!12}
    \cellcolor{white!20} \multirow{-5}{*} {\centering {TQA}} & \multirow{-3}{*} {\centering  Progressive Retrieval} & 600K $\rightarrow$ 80c & 80c $\xrightarrow[]{\sim \text{800d}}$ 12d & 12d $\xrightarrow[]{\sim \text{65p}}$ 2p & 3.47 (0.00+2.52+0.95) & 78.31 \\

\bottomrule
\end{tabular}
\caption{Retrieval performance comparison (in Top-2) \textit{w.r.t.} time cost and answer recall on NQ and TQA datasets.}
\label{table:nq_tqa_retrieval_performance_top2}
\end{table*}
\begin{table*}[!ht]
\centering
\footnotesize
\tabcolsep=0.125cm
\renewcommand\arraystretch{1.25}
\begin{tabular}{ccllllc}
\toprule
    \multirow{2}{*}[-0.5ex]{\textbf{Datasets}} &
    \multirow{2}{*}[-0.5ex]{\textbf{Retrieval Paradigm}} &
    \multicolumn{3}{c}{\textbf{Retrieval Stages}} &
    \multicolumn{1}{c}{\multirow{2}{*}[-0.4ex]{\textbf{\begin{tabular}[c]{@{}c@{}}Time Cost \\ (T)\end{tabular}}}} &
    \multirow{2}{*}[-0.4ex]{\textbf{\begin{tabular}[c]{@{}c@{}}Answer Recall \\ (AR)\end{tabular}}} \\ 
    \cmidrule(lr){3-5}
    &  &  \multicolumn{1}{c}{\centering Retrieval} & \multicolumn{1}{c}{Pre/Re-ranking} & \multicolumn{1}{c}{Post-ranking} &  &  \\ 
    \midrule
    & \multirow{2}{*}{Flat Retrieval}  &   21M $\rightarrow$ 3p  &  N/A  &  N/A   &  4.90 (4.90+N/A+N/A) & 69.64 \\
    & & 21M $\rightarrow$ 400p & 400p $\rightarrow$ 3p &  N/A &  5.25 (4.90+0.35+N/A) & 72.83 \\
    \rowcolor{gray!12}
    \cellcolor{white!20} &  & 600K $\rightarrow$ 3c & N/A  &  N/A  &  0.00 (0.00+N/A+N/A) &  68.98 \\
    \rowcolor{gray!12}
   \cellcolor{white!20} &  & 600K $\rightarrow$ 20c & 20c $\xrightarrow[]{\sim \text{180d}}$ 3d  &  N/A  &  0.49 (0.00+0.49+N/A) & 72.71   \\
    \rowcolor{gray!12}
    \cellcolor{white!20} \multirow{-5}{*}{\centering {NQ}} & \multirow{-3}{*}[0.5ex]{Progressive Retrieval} & 600K $\rightarrow$ 80c & 80c $\xrightarrow[]{\sim \text{740d}}$ 8d & 8d $\xrightarrow[]{\sim \text{50p}}$ 3p & 2.97 (0.00+2.20+0.77) & 73.88 \\
    \midrule
     & \multirow{2}{*}{Flat Retrieval}  &   21M $\rightarrow$ 3p  &  N/A  &  N/A   &  5.02 (5.02+N/A+N/A) & 73.11  \\
    &  & 21M $\rightarrow$ 400p  &  400p $\rightarrow$ 3p  &  N/A   &  5.41 (5.02+0.39+N/A) & 79.88 \\
    \rowcolor{gray!12}
    \cellcolor{white!20} &  & 600K $\rightarrow$ 3c & N/A  &  N/A  &  0.00 (0.00+N/A+N/A) &  76.96 \\
    \rowcolor{gray!12}
    \cellcolor{white!20} &  & 600K $\rightarrow$ 20c & 20c $\xrightarrow[]{\sim \text{200d}}$ 3d  &  N/A  &  0.60 (0.00+0.60+N/A) &  79.04  \\ 
    \rowcolor{gray!12}
    \cellcolor{white!20} \multirow{-5}{*} {\centering {TQA}} & \multirow{-3}{*} {\centering  Progressive Retrieval} & 600K $\rightarrow$ 80c & 80c $\xrightarrow[]{\sim \text{800d}}$ 12d & 12d $\xrightarrow[]{\sim \text{65p}}$ 3p & 3.47 (0.00+2.52+0.95) & 79.83 \\
\bottomrule
\end{tabular}
\caption{Retrieval performance comparison (in Top-3) \textit{w.r.t.} time cost and answer recall on NQ and TQA datasets.}
\label{table:nq_tqa_retrieval_performance_top3}
\end{table*}
\section{Document Clustering Algorithm}
\label{appen:cluster_document_algorithm}
Algorithm \ref{alg:cluster_documents} shows the document clustering algorithm used in the retrieval stage (\S \ref{sec:retrieval_stage}). Simply put, documents that are highly relevant will be clustered together, where each cluster is a list of documents related to each other. And, the clustered documents serve as the coarse-grained retrieval units.

\end{document}